\documentclass[a4paper,noinfo,crop]{bioinfo}
\usepackage{pdfpages}
\copyrightyear{2016}
\pubyear{2016}

\access{Advance Access Publication Date: Day Month Year}
\appnotes{Applications Note}

\usepackage{url}
\usepackage{tikz}

\makeatletter
\newcommand{\removelatexerror}{\let\@latex@error\@gobble}
\makeatother

\newcommand{\q}{\phantom0}
\newcommand{\qq}{\q\q}

\newcommand{\qqc}{\phantom{00,}}

\newcommand{\qqqqqc}{\phantom{00,000}}

\newcommand{\OoT}{\textit{Out of time ($>$12h)}}

\newcommand{\UnsK}{\textit{unsupported k}}

\newlength{\vnalen}
\newcommand{\vna}[1]{{%
\settowidth{\vnalen}{#1}
\makebox[\the\vnalen][c]{---}%
}}

\newif\iffigsinpdf

\begin{document}
\firstpage{1}

\subtitle{Sequence analysis}

\title[KMC 3: counting and manipulating \emph{k}-mer statistics]{KMC 3: counting and manipulating \emph{k}-mer statistics}

\author[Kokot, D{\l}ugosz and Deorowicz]{%
Marek Kokot\,$^{\text{\sfb 1}}$,
Maciej D{\l}ugosz\,$^{\text{\sfb 1}}$ and 
Sebastian Deorowicz\,$^{\text{\sfb 1,}*}$}
\address{$^{1}$Institute of Informatics, Silesian University of Technology, Akademicka 16, 44-100 Gliwice, Poland}

\corresp{$^\ast$To whom correspondence should be addressed.}

\history{Received on XXXXX; revised on XXXXX; accepted on XXXXX}

\editor{Associate Editor: XXXXXXX}

\abstract{\textbf{Summary:}
Counting all $k$-mers in a given dataset is a standard procedure in many bioinformatics applications.
We introduce KMC3, a significant improvement of the former KMC2 algorithm together with KMC tools for manipulating $k$-mer databases.
Usefulness of the tools is shown on a few real problems.
\\
\textbf{Availability:} Program is freely available at \url{http://sun.aei.polsl.pl/REFRESH/kmc}.\\
\textbf{Contact:} \href{sebastian.deorowicz@polsl.pl}{sebastian.deorowicz@polsl.pl}\\
}

\maketitle

\section{Introduction}
In many applications related to genome sequencing, e.g., \emph{de novo} assembly, read correction, repeat detection, comparison of genomes, the first step is \emph{$k$-mer counting}.
This procedure consists in determining all unique $k$-symbol long strings (usually with counters)  in the read collection.

From the conceptual point of view $k$-mer counting is quite simple task.
Nevertheless, the problems appear when we deal with real data, which could be huge.
There are many articles published in the recent years that discuss this problem, which shows that providing an efficient tool solving this task is far from trivial.

The obtained $k$-mer statistics are of course only a point of departure for following analyzes.
In some of them various operations on sets of $k$-mers are necessary.
We introduce a new version of KMC2~\citep{DKGD15}, which 
is more memory frugal and much faster.
We also introduce KMC tools for easy manipulation of sets produced by KMC.
The usefulness of KMC tools is shown on a few literature case studies in which various utilities were replaced by operations from our package with significant gains in processing time and memory usage.


\begin{methods}
\section{Methods}
KMC3 follows the same two-stage processing scheme as KMC2.
In the first stage the reads are split into several hundred bins (disk files) according to the signatures (short $m$-symbol long substrings) of $k$-mers.
The bins are then sorted one be one to remove duplicates in the second stage.
A~similar processing was used in several recently published algorithms for $k$-mer counting.
Their first stages are similar to KMC2, but then the bins are handled in different ways, e.g., Gerbil~\citep{ERM2016} employs hash tables, DSK2~\citep{RLC13} uses multistage processing in a case of small disk space.
Recently published, KCMBT~\citep{MPR2016} applies burst tries for $k$-mer storage.

\begin{figure}[!b]
\centering\includegraphics[width=0.85\columnwidth]{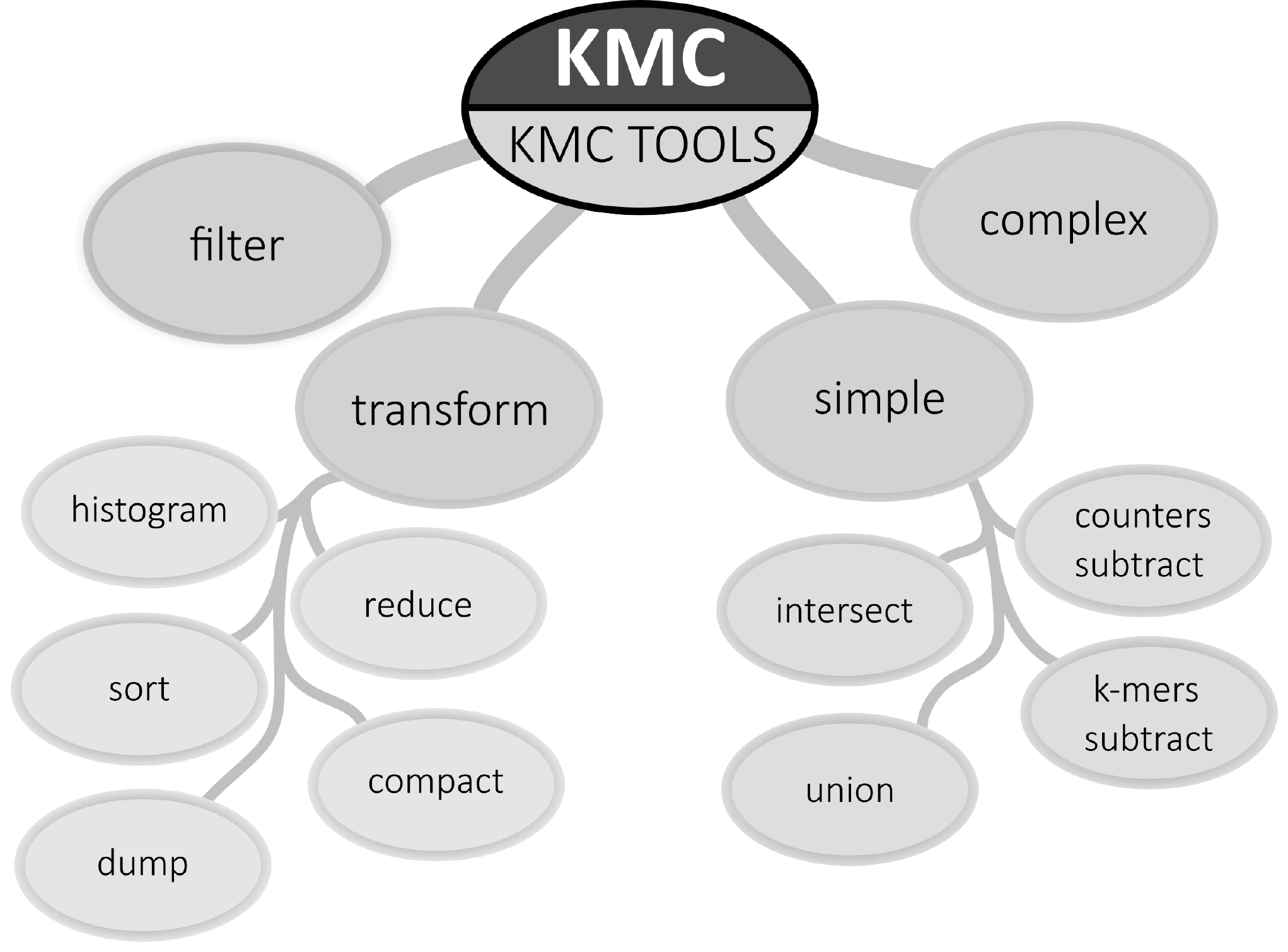}
\caption{Scheme of KMC 3 package}
\label{fig:tools}
\end{figure}

There are several main novelties in KMC3.
Concerning the first stage: input files, especially in gzipped FASTQ format, are loaded faster due to better input/output (I/O) subsystem, signatures are assigned to bins in an improved way, which results in smaller memory requirements.
The most significant improvements are in the second stage.
We replaced the radix sort used in KMC~\citep{DDG13} and KMC2 by our algorithm~\citep{KDG2016}.
We also improved the parallelization scheme of other routines (not directly related to sorting) at this stage.

The second part of the package consists of various tools for manipulation of sets containing $k$-mers.
Figure~\ref{fig:tools} shows a general scheme of the package (complete description is given in the Supplementary material).
The \emph{filtering} allows to extract from FASTQ files the reads satisfying some criteria, e.g., with sufficiently large number of $k$-mers occurrences.
The family of \emph{transform} operations allows to modify the KMC database, e.g., remove too frequent/rare $k$-mers, remove counters, build histogram.
The family of \emph{simple} operations is composed of several set operations involving more than one KMC database.
E.g., the user is able to intersect, union the databases.
Finally, the \emph{complex} operations allow to construct compound expressions taking as parameters KMC databases.


\section{Results}
We used six datasets for KMC3 evaluation.
Table~\ref{tab:results:counting} shows the results for two representatives.
The description of the datasets, the platform used for tests, the remaining results are given in the Supplementary material.
FASTQ files are almost always stored in compressed form.
Thus, we provide the running times for both: uncompressed and gziped FASTQ.

As we can observe for the \emph{G.~gallus} data, KMC3 is usually the fastest, especially for larger~$k$.
The times for \emph{H.~sapiens}~3, the largest of our dataset, are much longer.
Nevertheless, KMC3 was able to complete in less than 100 minutes for typical input format for both examined~$k$.
It can be observed, that the improved I/O subsystem 
as well as the new sorting routine and better parallelization of the second stage 
gave substantial benefits comparing to KMC2.
It is also worth noting that KMC3 even for the largest dataset used a reasonable amount of memory.

\begin{table}[t]
\processtable{Comparison of $k$-mers counting algorithms
\label{tab:results:counting}}
{
\renewcommand{\tabcolsep}{0.3em}
\begin{tabular}{lcccccccc}
\toprule
Algorithm &&	\multicolumn{3}{c}{$k=28$}&&\multicolumn{3}{c}{$k=55$}\\\cline{3-5}\cline{7-9}
					&& RAM	& Disk	& Time/gz-Time	&& RAM	& Disk	& Time/gz-Time \\
\midrule
\multicolumn{9}{c}{\emph{\bfseries G.~gallus} (35\,Gbases in total)}\\
DSK 2				&&	\q14	&	\q39 & \qqc 783 / \q1,180	&& 15	& 30	& \qqc719 / \q1,123 	\\
Gerbil			&& \qq2	&\q24	& \qqc607 / \qqc631	 		&& \q4	&\q 17	& \qqc615 / \qqc551 	\\
GTester4			&& \q74	& \qq0	& \q2,494 / \vna{\qqqqqc} && \multicolumn{3}{c}{\UnsK} \\
Jellyfish 2		&&	\q33	& \qq0	& \qqc909 / \vna{\qqqqqc}	&& 77		&\qq0 	& \q1,048 / \vna{\qqqqqc}\\
KCMBT			 	&&	107	& \qq0	& \q1,335 / \vna{\qqqqqc}	&& \multicolumn{3}{c}{\UnsK} \\
KMC 2				&& \q12		&\q25	&\qqc 489 /\qqc321 		&& 12		&\q 18	& \qqc861 / \qqc672  	\\
KMC 3				&&	\q12		&\q28	&\qqc 492 /\qqc292 		&&	12		&\qq 18	& \qqc455 / \qqc245 	\\
\midrule
\multicolumn{9}{c}{\emph{\bfseries H.~sapiens}~3 (729\,Gbases in total)}\\
Gerbil			&& \q28		& 523	& 11,994 / 12,730		&& 62	& 364	& 11,968	/ 12,469 \\
Jellyfish 2		&&	121	& 251	& 43,005 / \vna{\qqqqqc}	&& \multicolumn{3}{c}{\OoT} \\
KMC 2				&& \q64		& 551	& 10,777 /	\q 9,036	&& 72	& 381	& 13,774 / 11,804 \\
KMC 3				&&	\q33		& 596	& \q9,631 / \q 5,985		&& 34	& 389	& \q8,750 / \q5,331 \\
\botrule
\end{tabular}
}{The times are given for: uncompressed input FASTQ file (`Time') and gzipped input FASTQ files (`gz-Time')
The units are: seconds (time), GB (Disk and RAM).
All programs were executed in 12-threads mode, except for KCMBT which was executed in 8-threads mode due to large main memory consumption (exceeding 128\,GB RAM for 12 threads). 
`---' means that the mode is not supported.
There are no results for some programs for \emph{H.~sapiens}~3 due to: processing longer than 12h (DSK 2, GTester4) or memory requirements larger than 128\,GB (KCMBT).
}
\end{table}

To evaluate KMC tools we picked three studies described recently in the literature.
Below we briefly describe the goals of the studies and total gains in time and memory.
The detailed results together with the scripts showing KMC tools usage are given in the Supplementary material.

DIAMUND~\citep{SPFS2014} is a novel approach for variant detection.
It is dedicated to comparison of family trios or normal and diseased samples of the same individual.
Instead of mapping the reads onto a reference genome, DIAMUND compares directly the raw reads.
We followed this protocol for the Ashkenazim Jewish ancestry trio~\citep{ZCD2016}.
DIAMUND needed 13 hours to complete its work and used 107\,GB RAM.
As most of the stages can be made using KMC tools we replaced the original tools used in DIAMUND and reduced the processing time to 4 hours (reducing memory usage to 12\,GB RAM).

NIKS~\citep{NAJ2013} is another tool that uses raw sequencing reads for mutation identification taking into account mainly $k$-mer statistics.
We replaced its stages related to $k$-mer counting and manipulation of $k$-mer databases by KMC tools operations.
In a single stage it was necessary to use KMC API to prepare a short C++ program to reproduce the format of intermediary data used in NIKS.
The processing time was reduced from approx.\ 40 minutes to 5 minutes in this case (RAM usage was reduced from 92\,GB to 12\,GB).

Finally, we made the same experiment as used in the GenomeTester4 article~\citep{ERM2016}, introducing a tool that is able to perform similar operations as KMC tools.
The investigated problem was to find group-specific $k$-mers to identify bacteria.
The processing time was reduced from 5 hours to 15 minutes (RAM usage reduction from 75\,GB to 12\,GB).


\end{methods}
\section{Discussion}
We proposed a new version of our $k$-mer counter. 
It is much faster (even a few times for large $k$) than its predecessor and faster than the existing competitors.
The KMC tools offer a number of operations on $k$-mer databases that can be used in projects making use of $k$-mer statistics.



\section*{Funding}
This work was supported by the Polish National Science Centre under the project DEC-2015/17/B/ST6/01890 and
by Silesian University of Technology grant no.\ BKM507/RAU2/2016.
We used the infrastructure supported by POIG.02.03.01-24-099/13 grant:
``GeCONiI--Upper Silesian Center for Computational Science and
Engineering''.


\clearpage


\section*{Supplementary material}

# KMC 3: counting and manipulating $k$-mer statistics
# Supplementary material

Marek Kokot, Maciej Długosz, and Sebastian Deorowicz

January 27, 2017

# Contents

# 1 Examined programs

## 1.1 General parameters of programs

The following programs were used in the experimental part. The running parameters are also given.

- Jellyfish v. 2.2.6
    - `count -m <kmer_size> -C -s <hash_size> -t <threads> -L 2 -o <output> <input>`

    hash_size value was dependent on the input sequence size

- DSK v. 2.1.0
    - `-file <input> -kmer-size <kmer_size> -verbose 2 -nb-cores <threads> -out <output> -abundance-min 2`

- KCMBT v. 1.0.0
    - `-k <kmer_size> -t <threads> -i <input>`

    maximum supported value of kmer_size is 32, thus we run KCMBT only for kmer_size = 28

- Gerbil v. 0.2
    - `-k <kmer_size> -l 2 -t <threads> -i <input> tmp/ <output>`

- glistmaker v. 4.0
    - `<input> -w <kmer_size> -c 2 --num_threads <threads> -o <output>`

- KMC v. 2.3.0
    - `-k<kmer_size> -ci2 -t<threads> -v <input> <output> tmp`

- KMC v. 3.0.0
    - `-k<kmer_size> -ci2 -t<threads> -v <input> <output> tmp`

## 1.2 Exact command line parameters

In this section we present exact parameters used in our experiments. This command lines was used on XEON machine, yet the only difference between parameters on XEON and Opteron is the number of thread which was set to 16 in case of the latter one. For gzipped input list of input files is stored in a text file *.txt. For example hs1.txt contains list of gzipped FASTQ files of *H. sapiens* 1 dataset.

**DSK2**

Command lines used for input in FASTQ format:

```
./dsk -file fv.fastq -kmer-size 28 -max-memory 12000 -verbose 2 -nb-cores 12 -out res
-abundance-min 2
./dsk -file fv.fastq -kmer-size 40 -max-memory 12000 -verbose 2 -nb-cores 12 -out res
-abundance-min 2
./dsk -file fv.fastq -kmer-size 55 -max-memory 12000 -verbose 2 -nb-cores 12 -out res
-abundance-min 2
./dsk -file fv.fastq -kmer-size 65 -max-memory 12000 -verbose 2 -nb-cores 12 -out res
-abundance-min 2

./dsk -file gg.fastq -kmer-size 28 -max-memory 12000 -verbose 2 -nb-cores 12 -out res
-abundance-min 2
```

```
./dsk -file gg.fastq -kmer-size 40 -max-memory 12000 -verbose 2 -nb-cores 12 -out res
-abundance-min 2
./dsk -file gg.fastq -kmer-size 55 -max-memory 12000 -verbose 2 -nb-cores 12 -out res
-abundance-min 2
./dsk -file gg.fastq -kmer-size 65 -max-memory 12000 -verbose 2 -nb-cores 12 -out res
-abundance-min 2

./dsk -file mb.fastq -kmer-size 28 -max-memory 12000 -verbose 2 -nb-cores 12 -out res
-abundance-min 2
./dsk -file mb.fastq -kmer-size 40 -max-memory 12000 -verbose 2 -nb-cores 12 -out res
-abundance-min 2
./dsk -file mb.fastq -kmer-size 55 -max-memory 12000 -verbose 2 -nb-cores 12 -out res
-abundance-min 2
./dsk -file mb.fastq -kmer-size 65 -max-memory 12000 -verbose 2 -nb-cores 12 -out res
-abundance-min 2

./dsk -file hs1.fastq -kmer-size 28 -max-memory 12000 -verbose 2 -nb-cores 12 -out res
-abundance-min 2
./dsk -file hs1.fastq -kmer-size 40 -max-memory 12000 -verbose 2 -nb-cores 12 -out res
-abundance-min 2
./dsk -file hs1.fastq -kmer-size 55 -max-memory 12000 -verbose 2 -nb-cores 12 -out res
-abundance-min 2
./dsk -file hs1.fastq -kmer-size 65 -max-memory 12000 -verbose 2 -nb-cores 12 -out res
-abundance-min 2

./dsk -file hs2.fastq -kmer-size 28 -max-memory 12000 -verbose 2 -nb-cores 12 -out res
-abundance-min 2
./dsk -file hs2.fastq -kmer-size 40 -max-memory 12000 -verbose 2 -nb-cores 12 -out res
-abundance-min 2
./dsk -file hs2.fastq -kmer-size 55 -max-memory 12000 -verbose 2 -nb-cores 12 -out res
-abundance-min 2
./dsk -file hs2.fastq -kmer-size 65 -max-memory 12000 -verbose 2 -nb-cores 12 -out res
-abundance-min 2

./dsk -file hs3.fastq -kmer-size 28 -max-memory 32000 -max-disk 600000 -verbose 2
-nb-cores 12 -out res -abundance-min 2
./dsk -file hs3.fastq -kmer-size 40 -max-memory 32000 -max-disk 480000 -verbose 2
-nb-cores 12 -out res -abundance-min 2
./dsk -file hs3.fastq -kmer-size 55 -max-memory 32000 -max-disk 390000 -verbose 2
-nb-cores 12 -out res -abundance-min 2
./dsk -file hs3.fastq -kmer-size 65 -max-memory 32000 -max-disk 350000 -verbose 2
-nb-cores 12 -out res -abundance-min 2
```

Command lines used for gzipped files:

```
./dsk -file fv.txt -kmer-size 28 -max-memory 12000 -verbose 2 -nb-cores 12 -out res
-abundance-min 2
./dsk -file fv.txt -kmer-size 40 -max-memory 12000 -verbose 2 -nb-cores 12 -out res
-abundance-min 2
./dsk -file fv.txt -kmer-size 65 -max-memory 12000 -verbose 2 -nb-cores 12 -out res
-abundance-min 2
./dsk -file fv.txt -kmer-size 55 -max-memory 12000 -verbose 2 -nb-cores 12 -out res
-abundance-min 2

./dsk -file gg.txt -kmer-size 28 -max-memory 12000 -verbose 2 -nb-cores 12 -out res
```

```
-abundance-min 2
./dsk -file gg.txt -kmer-size 40 -max-memory 12000 -verbose 2 -nb-cores 12 -out res
-abundance-min 2
./dsk -file gg.txt -kmer-size 55 -max-memory 12000 -verbose 2 -nb-cores 12 -out res
-abundance-min 2
./dsk -file gg.txt -kmer-size 65 -max-memory 12000 -verbose 2 -nb-cores 12 -out res
-abundance-min 2

./dsk -file mb.txt -kmer-size 28 -max-memory 12000 -verbose 2 -nb-cores 12 -out res
-abundance-min 2
./dsk -file mb.txt -kmer-size 40 -max-memory 12000 -verbose 2 -nb-cores 12 -out res
-abundance-min 2
./dsk -file mb.txt -kmer-size 55 -max-memory 12000 -verbose 2 -nb-cores 12 -out res
-abundance-min 2
./dsk -file mb.txt -kmer-size 65 -max-memory 12000 -verbose 2 -nb-cores 12 -out res
-abundance-min 2

./dsk -file hs1.txt -kmer-size 28 -max-memory 12000 -verbose 2 -nb-cores 12 -out res
-abundance-min 2
./dsk -file hs1.txt -kmer-size 40 -max-memory 12000 -verbose 2 -nb-cores 12 -out res
-abundance-min 2
./dsk -file hs1.txt -kmer-size 55 -max-memory 12000 -verbose 2 -nb-cores 12 -out res
-abundance-min 2
./dsk -file hs1.txt -kmer-size 65 -max-memory 12000 -verbose 2 -nb-cores 12 -out res
-abundance-min 2

./dsk -file hs2.txt -kmer-size 28 -max-memory 12000 -verbose 2 -nb-cores 12 -out res
-abundance-min 2
./dsk -file hs2.txt -kmer-size 40 -max-memory 12000 -verbose 2 -nb-cores 12 -out res
-abundance-min 2
./dsk -file hs2.txt -kmer-size 55 -max-memory 12000 -verbose 2 -nb-cores 12 -out res
-abundance-min 2
./dsk -file hs2.txt -kmer-size 65 -max-memory 12000 -verbose 2 -nb-cores 12 -out res
-abundance-min 2

./dsk -file hs3.txt -kmer-size 28 -max-memory 32000 -max-disk 600000 -verbose 2
-nb-cores 12 -out res -abundance-min 2
./dsk -file hs3.txt -kmer-size 40 -max-memory 32000 -max-disk 480000 -verbose 2
-nb-cores 12 -out res -abundance-min 2
./dsk -file hs3.txt -kmer-size 55 -max-memory 32000 -max-disk 390000 -verbose 2
-nb-cores 12 -out res -abundance-min 2
./dsk -file hs3.txt -kmer-size 65 -max-memory 32000 -max-disk 350000 -verbose 2
-nb-cores 12 -out res -abundance-min 2
```

On Opetron machine `-nb-cores 16` was used instead of `-nb-cores 12`

### Gerbil

Command lines used for input in FASTQ format:

```
./Gerbil -k 28 -l 2 -t 12 -i fv.fastq gbins/ o_ger
./Gerbil -k 40 -l 2 -t 12 -i fv.fastq gbins/ o_ger
./Gerbil -k 55 -l 2 -t 12 -i fv.fastq gbins/ o_ger
./Gerbil -k 65 -l 2 -t 12 -i fv.fastq gbins/ o_ger
```

```
./Gerbil -k 28 -l 2 -t 12 -i gg.fastq gbins/ o_ger
./Gerbil -k 40 -l 2 -t 12 -i gg.fastq gbins/ o_ger
./Gerbil -k 55 -l 2 -t 12 -i gg.fastq gbins/ o_ger
./Gerbil -k 65 -l 2 -t 12 -i gg.fastq gbins/ o_ger

./Gerbil -k 28 -l 2 -t 12 -i mb.fastq gbins/ o_ger
./Gerbil -k 40 -l 2 -t 12 -i mb.fastq gbins/ o_ger
./Gerbil -k 55 -l 2 -t 12 -i mb.fastq gbins/ o_ger
./Gerbil -k 65 -l 2 -t 12 -i mb.fastq gbins/ o_ger

./Gerbil -k 28 -l 2 -t 12 -i hs1.fastq gbins/ res
./Gerbil -k 40 -l 2 -t 12 -i hs1.fastq gbins/ res
./Gerbil -k 55 -l 2 -t 12 -i hs1.fastq gbins/ res
./Gerbil -k 65 -l 2 -t 12 -i hs1.fastq gbins/ res

./Gerbil -k 28 -l 2 -t 12 -i hs2.fastq gbins/ o_ger
./Gerbil -k 40 -l 2 -t 12 -i hs2.fastq gbins/ o_ger
./Gerbil -k 55 -l 2 -t 12 -i hs2.fastq gbins/ o_ger
./Gerbil -k 65 -l 2 -t 12 -i hs2.fastq gbins/ o_ger

./Gerbil -k 28 -l 2 -t 12 -i hs3.fastq gbins/ res
./Gerbil -k 40 -l 2 -t 12 -i hs3.fastq gbins/ res
./Gerbil -k 55 -l 2 -t 12 -i hs3.fastq gbins/ res
./Gerbil -k 65 -l 2 -t 12 -i hs3.fastq gbins/ res
```

Command lines used for gzipped files:

```
./Gerbil -k 28 -l 2 -t 12 -i fv.txt gbins/ o_ger
./Gerbil -k 40 -l 2 -t 12 -i fv.txt gbins/ o_ger
./Gerbil -k 55 -l 2 -t 12 -i fv.txt gbins/ o_ger
./Gerbil -k 65 -l 2 -t 12 -i fv.txt gbins/ o_ger

./Gerbil -k 28 -l 2 -t 12 -i gg.txt gbins/ res
./Gerbil -k 40 -l 2 -t 12 -i gg.txt gbins/ res
./Gerbil -k 55 -l 2 -t 12 -i gg.txt gbins/ res
./Gerbil -k 65 -l 2 -t 12 -i gg.txt gbins/ res

./Gerbil -k 28 -l 2 -t 12 -i mb.txt gbins/ res
./Gerbil -k 40 -l 2 -t 12 -i mb.txt gbins/ res
./Gerbil -k 55 -l 2 -t 12 -i mb.txt gbins/ res
./Gerbil -k 65 -l 2 -t 12 -i mb.txt gbins/ res

./Gerbil -k 28 -l 2 -t 12 -i hs1.txt gbins/ res
./Gerbil -k 40 -l 2 -t 12 -i hs1.txt gbins/ res
./Gerbil -k 55 -l 2 -t 12 -i hs1.txt gbins/ res
./Gerbil -k 65 -l 2 -t 12 -i hs1.txt gbins/ res

./Gerbil -k 28 -l 2 -t 12 -i hs2.txt gbins/ o_ger
./Gerbil -k 40 -l 2 -t 12 -i hs2.txt gbins/ res
./Gerbil -k 55 -l 2 -t 12 -i hs2.txt gbins/ res
./Gerbil -k 65 -l 2 -t 12 -i hs2.txt gbins/ res

./Gerbil -k 28 -l 2 -t 12 -i hs3.txt gbins/ res
./Gerbil -k 40 -l 2 -t 12 -i hs3.txt gbins/ res
./Gerbil -k 55 -l 2 -t 12 -i hs3.txt gbins/ res
```

```
./Gerbil -k 65 -l 2 -t 12 -i hs3.txt gbins/ res
```

On Opetron machine `-t 16` was used instead of `-t 12`

**GenomeTester4**

GenomeTester4 supports only input files in FASTQ format and $k$ below 33.

```
./glistmaker fv.fastq -w 28 --num_threads 12 -c 2 -o res
./glistmaker gg.fastq -w 28 --num_threads 12 -c 2 -o res
./glistmaker mb.fastq -w 28 --num_threads 12 -c 2 -o res
./glistmaker hs1.fastq -w 28 --num_threads 12 -c 2 -o gres
./glistmaker hs2.fastq -w 28 --num_threads 12 -c 2 -o res
./glistmaker hs3.fastq -w 28 --num_threads 12 -c 2 -o gres
```

On Opetron machine `--num_threads 16` was used instead of `--num_threads 12`

**Jellyfish**

Jellyfish supports only input files in FASTQ format.

```
./jellyfish count -m 28 -C -s 300M -t 12 -L 2 -o jelly2 fv.fastq
./jellyfish count -m 40 -C -s 400M -t 12 -L 2 -o jelly2 fv.fastq
./jellyfish count -m 55 -C -s 400M -t 12 -L 2 -o jelly2 fv.fastq
./jellyfish count -m 65 -C -s 400M -t 12 -L 2 -o jelly2 fv.fastq

./jellyfish count -m 28 -C -s 1200M -t 12 -L 2 -o jelly2 gg.fastq
./jellyfish count -m 40 -C -s 1200M -t 12 -L 2 -o jelly2 gg.fastq
./jellyfish count -m 55 -C -s 1200M -t 12 -L 2 -o jelly2 gg.fastq
./jellyfish count -m 65 -C -s 1200M -t 12 -L 2 -o jelly2 gg.fastq

./jellyfish count -m 28 -C -s 1G -t 12 -L 2 -o jelly2 mb.fastq
./jellyfish count -m 40 -C -s 1200M -t 12 -L 2 -o jelly2 mb.fastq
./jellyfish count -m 55 -C -s 1200M -t 12 -L 2 -o jelly2 mb.fastq
./jellyfish count -m 65 -C -s 1200M -t 12 -L 2 -o jelly2 mb.fastq

./jellyfish count -m 28 -C -s 3G -t 12 -L 2 -o jelly2 hs1.fastq
./jellyfish count -m 40 -C -s 3G -t 12 -L 2 -o jelly2 hs1.fastq
./jellyfish count -m 55 -C -s 3G -t 12 -L 2 -o jelly2 hs1.fastq
./jellyfish count -m 65 -C -s 3G -t 12 -L 2 -o jelly2 hs1.fastq

./jellyfish count -m 28 -C -s 3G -t 12 -L 2 -o jelly2 hs2.fastq
./jellyfish count -m 40 -C -s 3G -t 12 -L 2 -o jelly2 hs2.fastq
./jellyfish count -m 55 -C -s 3G -t 12 -L 2 -o jelly2 hs2.fastq
./jellyfish count -m 65 -C -s 3G -t 12 -L 2 -o jelly2 hs2.fastq

./jellyfish count -m 28 -C -s 3G -t 12 -L 2 -o res hs3.fastq
./jellyfish count -m 40 -C -s 3G -t 12 -L 2 -o res hs3.fastq
./jellyfish count -m 55 -C -s 3G -t 12 -L 2 -o res hs3.fastq
./jellyfish count -m 65 -C -s 3G -t 12 -L 2 -o res hs3.fastq
```

On Opetron machine `-t 16` was used instead of `-t 12`

**KCMBT**

KCMBT supports only input files in FASTQ format and $k$ below 33. The documentation says that a number of threads should be a power of two so on XEON machine thread number was set to 8.

```
./kcmbt -k 28 -t 8 -i fv.fastq
./kcmbt -k 28 -t 8 -i gg.fastq
./kcmbt -k 28 -t 8 -i mb.fastq
./kcmbt -k 28 -t 8 -i hs1.fastq
./kcmbt -k 28 -t 8 -i hs2.fastq
./kcmbt -k 28 -t 8 -i hs3.fastq
```

On Opetron machine `-t 16` was used instead of `-t 8`

**KMC2**

Command lines used for input in FASTQ format:

```
./kmc2 -v -t12 -m12 -k28 fv.fastq res ./bins
./kmc2 -v -t12 -m12 -k40 fv.fastq res ./bins
./kmc2 -v -t12 -m12 -k55 fv.fastq res ./bins
./kmc2 -v -t12 -m12 -k65 fv.fastq res ./bins

./kmc2 -v -t12 -m12 -k28 gg.fastq res ./bins
./kmc2 -v -t12 -m12 -k40 gg.fastq res ./bins
./kmc2 -v -t12 -m12 -k55 gg.fastq res ./bins
./kmc2 -v -t12 -m12 -k65 gg.fastq res ./bins

./kmc2 -v -t12 -m12 -k28 mb.fastq res ./bins
./kmc2 -v -t12 -m12 -k40 mb.fastq res ./bins
./kmc2 -v -t12 -m12 -k55 mb.fastq res ./bins
./kmc2 -v -t12 -m12 -k65 mb.fastq res ./bins

./kmc2 -ci2 -v -t12 -k28 hs1.fastq res ./bins
./kmc2 -ci2 -v -t12 -k40 hs1.fastq res ./bins
./kmc2 -ci2 -v -t12 -k55 hs1.fastq res ./bins
./kmc2 -ci2 -v -t12 -k65 hs1.fastq res ./bins

./kmc2 -v -t12 -m12 -k28 hs2.fastq res ./bins
./kmc2 -v -t12 -m12 -k40 hs2.fastq res ./bins
./kmc2 -v -t12 -m12 -k55 hs2.fastq res ./bins
./kmc2 -v -t12 -m12 -k65 hs2.fastq res ./bins

./kmc2 -ci2 -v -t12 -m64 -k28 hs3.fastq res ./bins
./kmc2 -ci2 -v -t12 -m64 -k40 hs3.fastq res ./bins
./kmc2 -ci2 -v -t12 -m64 -k55 hs3.fastq res ./bins
./kmc2 -ci2 -v -t12 -m64 -k65 hs3.fastq res ./bins
```

Command lines used for gzipped files:

```
./kmc2 -ci2 -v -t12 -k28 @fv.txt res ./bins
./kmc2 -ci2 -v -t12 -k40 @fv.txt res ./bins
./kmc2 -ci2 -v -t12 -k55 @fv.txt res ./bins
./kmc2 -ci2 -v -t12 -k65 @fv.txt res ./bins

./kmc2 -ci2 -v -t12 -k28 @gg.txt res ./bins
./kmc2 -ci2 -v -t12 -k40 @gg.txt res ./bins
./kmc2 -ci2 -v -t12 -k55 @gg.txt res ./bins
./kmc2 -ci2 -v -t12 -k65 @gg.txt res ./bins

./kmc2 -ci2 -v -t12 -k28 @mb.txt res ./bins
```

7

```
./kmc2 -ci2 -v -t12 -k40 @mb.txt res ./bins
./kmc2 -ci2 -v -t12 -k55 @mb.txt res ./bins
./kmc2 -ci2 -v -t12 -k65 @mb.txt res ./bins

./kmc2 -ci2 -v -t12 -k28 @hs1.txt res ./bins
./kmc2 -ci2 -v -t12 -k40 @hs1.txt res ./bins
./kmc2 -ci2 -v -t12 -k55 @hs1.txt res ./bins
./kmc2 -ci2 -v -t12 -k65 @hs1.txt res ./bins

./kmc2 -ci2 -v -t12 -k28 @hs2.txt res ./bins
./kmc2 -ci2 -v -t12 -k40 @hs2.txt res ./bins
./kmc2 -ci2 -v -t12 -k55 @hs2.txt res ./bins
./kmc2 -ci2 -v -t12 -k65 @hs2.txt res ./bins

./kmc2 -ci2 -v -t12 -m64 -k28 @hs3.txt res ./bins
./kmc2 -ci2 -v -t12 -m64 -k40 @hs3.txt res ./bins
./kmc2 -ci2 -v -t12 -m64 -k55 @hs3.txt res ./bins
./kmc2 -ci2 -v -t12 -m64 -k65 @hs3.txt res ./bins
```

On Opetron machine `-t16` was used instead of `-t12`

**KMC3**

Command lines used for input in FASTQ format:

```
./kmc3 -ci2 -v -t12 -k28 fv.fastq res ./bins
./kmc3 -ci2 -v -t12 -k40 fv.fastq res ./bins
./kmc3 -ci2 -v -t12 -k55 fv.fastq res ./bins
./kmc3 -ci2 -v -t12 -k65 fv.fastq res ./bins

./kmc3 -ci2 -v -t12 -k28 gg.fastq res ./bins
./kmc3 -ci2 -v -t12 -k40 gg.fastq res ./bins
./kmc3 -ci2 -v -t12 -k55 gg.fastq res ./bins
./kmc3 -ci2 -v -t12 -k65 gg.fastq res ./bins

./kmc3 -ci2 -v -t12 -k28 mb.fastq res ./bins
./kmc3 -ci2 -v -t12 -k40 mb.fastq res ./bins
./kmc3 -ci2 -v -t12 -k55 mb.fastq res ./bins
./kmc3 -ci2 -v -t12 -k65 mb.fastq res ./bins


./kmc3 -ci2 -v -t12 -k28 hs1.fastq res ./bins
./kmc3 -ci2 -v -t12 -k40 hs1.fastq res ./bins
./kmc3 -ci2 -v -t12 -k55 hs1.fastq res ./bins
./kmc3 -ci2 -v -t12 -k65 hs1.fastq res ./bins

./kmc3 -ci2 -v -t12 -k28 hs2.fastq res ./bins
./kmc3 -ci2 -v -t12 -k40 hs2.fastq res ./bins
./kmc3 -ci2 -v -t12 -k55 hs2.fastq res ./bins
./kmc3 -ci2 -v -t12 -k65 hs2.fastq res ./bins

./kmc3 -ci2 -v -t12 -m32 -k28 hs3.fastq res ./bins
./kmc3 -ci2 -v -t12 -m32 -k40 hs3.fastq res ./bins
./kmc3 -ci2 -v -t12 -m32 -k55 hs3.fastq res ./bins
./kmc3 -ci2 -v -t12 -m32 -k65 hs3.fastq res ./bins
```

Command lines used for gzipped files:

```
./kmc3 -ci2 -v -t12 -k28 @fv.txt res ./bins
./kmc3 -ci2 -v -t12 -k40 @fv.txt res ./bins
./kmc3 -ci2 -v -t12 -k55 @fv.txt res ./bins
./kmc3 -ci2 -v -t12 -k65 @fv.txt res ./bins

./kmc3 -ci2 -v -t12 -k28 @gg.txt res ./bins
./kmc3 -ci2 -v -t12 -k40 @gg.txt res ./bins
./kmc3 -ci2 -v -t12 -k55 @gg.txt res ./bins
./kmc3 -ci2 -v -t12 -k65 @gg.txt res ./bins

./kmc3 -ci2 -v -t12 -k28 @mb.txt res ./bins
./kmc3 -ci2 -v -t12 -k40 @mb.txt res ./bins
./kmc3 -ci2 -v -t12 -k55 @mb.txt res ./bins
./kmc3 -ci2 -v -t12 -k65 @mb.txt res ./bins

./kmc3 -ci2 -v -t12 -k28 @hs1.txt res ./bins
./kmc3 -ci2 -v -t12 -k40 @hs1.txt res ./bins
./kmc3 -ci2 -v -t12 -k55 @hs1.txt res ./bins
./kmc3 -ci2 -v -t12 -k65 @hs1.txt res ./bins

./kmc3 -ci2 -v -t12 -k28 @hs2.txt res ./bins
./kmc3 -ci2 -v -t12 -k40 @hs2.txt res ./bins
./kmc3 -ci2 -v -t12 -k55 @hs2.txt res ./bins
./kmc3 -ci2 -v -t12 -k65 @hs2.txt res ./bins

./kmc3 -ci2 -v -t12 -m32 -k28 @hs3.txt res ./bins
./kmc3 -ci2 -v -t12 -m32 -k40 @hs3.txt res ./bins
./kmc3 -ci2 -v -t12 -m32 -k55 @hs3.txt res ./bins
./kmc3 -ci2 -v -t12 -m32 -k65 @hs3.txt res ./bins
```

On Opetron machine `-t16` was used instead of `-t12`

# 2 Environment

Two computers were used for performing the experiments:

- Intel Xeon-based workstation:
  - 2 Intel Xeon E5-2670 CPUs; 12 cores per CPU, each clocked at 2.3 GHz,
  - 128 GB RAM,
  - Samsung 850 EVO SSD of size 1 TB; `hdparm -t` reported speed: 430 MB/s,
  - 2 Seagate Enterprise NAS HDD of size 6 TB each in RAID0 configuration; `hdparm -t` reported speed: 360 MB/s.

- AMD Opteron-based server:
  - 4 AMD Opteron 6320 CPUs, 8 cores per CPU, each clocked at 2.8 GHz,
  - 256 GB RAM,
  - 5 Toshiba MG03SCA100 HDD of size 1 TB each in RAID5 configuration; `hdparm -t` reported speed: 540 MB/s.

# 3  $k$-mer counting

## 3.1  Datasets and running parameters

The datasets used in the experiments are characterized in Table 1. *M. balbisiana* files were downloaded in bzip2 format and recompressed to gzip format.

Table 1: Datasets used in the experiments. No. of bases are in Gbases. File sizes are in Gbytes (1Gbyte = $10^9$ bytes). Approximate genome lengths are in Mbases according to http://www.ncbi.nlm.nih.gov/genome/.

| Organism | Genome length | No. bases | FASTQ files size | No. files | Gzipped size | Avg. read length |
|---|---:|---:|---:|---:|---:|---:|
| *F. vesca* | 214 | 4.5 | 10.2 | 11 | 2.9 | 352 |
| *G. gallus* | 1,230 | 34.7 | 115.9 | 15 | 20.5 | 100 |
| *M. balbisiana* | 472 | 56.3 | 197.1 | 2 | 49.1 | 100 |
| *H. sapiens* 1 | 2,991 | 123.7 | 292.1 | 2 | 80.1 | 151 |
| *H. sapiens* 2 | 2,991 | 135.3 | 339.5 | 48 | 105.8 | 101 |
| *H. sapiens* 3 | 2,991 | 736.4 | 1,888.4 | 36 | 614.1 | 101 |

The files were downloaded from the following URLs:

- *F. vesca*
  ftp://ftp.ddbj.nig.ac.jp/ddbj_database/dra/fastq/SRA020/SRA020125/SRX030576/SRR072006.fastq.bz2
  ftp://ftp.ddbj.nig.ac.jp/ddbj_database/dra/fastq/SRA020/SRA020125/SRX030576/SRR072007.fastq.bz2
  ftp://ftp.ddbj.nig.ac.jp/ddbj_database/dra/fastq/SRA020/SRA020125/SRX030577/SRR072008.fastq.bz2
  ftp://ftp.ddbj.nig.ac.jp/ddbj_database/dra/fastq/SRA020/SRA020125/SRX030577/SRR072009.fastq.bz2
  ftp://ftp.ddbj.nig.ac.jp/ddbj_database/dra/fastq/SRA020/SRA020125/SRX030578/SRR072013.fastq.bz2
  ftp://ftp.ddbj.nig.ac.jp/ddbj_database/dra/fastq/SRA020/SRA020125/SRX030578/SRR072014.fastq.bz2
  ftp://ftp.ddbj.nig.ac.jp/ddbj_database/dra/fastq/SRA020/SRA020125/SRX030578/SRR072029.fastq.bz2
  ftp://ftp.ddbj.nig.ac.jp/ddbj_database/dra/fastq/SRA020/SRA020125/SRX030575/SRR072005.fastq.bz2
  ftp://ftp.ddbj.nig.ac.jp/ddbj_database/dra/fastq/SRA020/SRA020125/SRX030575/SRR072010.fastq.bz2
  ftp://ftp.ddbj.nig.ac.jp/ddbj_database/dra/fastq/SRA020/SRA020125/SRX030575/SRR072011.fastq.bz2
  ftp://ftp.ddbj.nig.ac.jp/ddbj_database/dra/fastq/SRA020/SRA020125/SRX030575/SRR072012.fastq.bz2

- *G. gallus*
  ftp://ftp.ddbj.nig.ac.jp/ddbj_database/dra/fastq/SRA030/SRA030308/SRX043656/SRR105788_1.fastq.bz2
  ftp://ftp.ddbj.nig.ac.jp/ddbj_database/dra/fastq/SRA030/SRA030308/SRX043656/SRR105788_2.fastq.bz2
  ftp://ftp.ddbj.nig.ac.jp/ddbj_database/dra/fastq/SRA030/SRA030309/SRX043656/SRR105789_1.fastq.bz2
  ftp://ftp.ddbj.nig.ac.jp/ddbj_database/dra/fastq/SRA030/SRA030309/SRX043656/SRR105789_2.fastq.bz2
  ftp://ftp.ddbj.nig.ac.jp/ddbj_database/dra/fastq/SRA030/SRA030312/SRX043656/SRR105792_1.fastq.bz2
  ftp://ftp.ddbj.nig.ac.jp/ddbj_database/dra/fastq/SRA030/SRA030312/SRX043656/SRR105792_2.fastq.bz2
  ftp://ftp.ddbj.nig.ac.jp/ddbj_database/dra/fastq/SRA030/SRA030314/SRX043656/SRR105794.fastq.bz2
  ftp://ftp.ddbj.nig.ac.jp/ddbj_database/dra/fastq/SRA030/SRA030314/SRX043656/SRR105794_1.fastq.bz2
  ftp://ftp.ddbj.nig.ac.jp/ddbj_database/dra/fastq/SRA030/SRA030314/SRX043656/SRR105794_2.fastq.bz2
  ftp://ftp.ddbj.nig.ac.jp/ddbj_database/dra/fastq/SRA036/SRA036382/SRX043656/SRR197985.fastq.bz2
  ftp://ftp.ddbj.nig.ac.jp/ddbj_database/dra/fastq/SRA036/SRA036382/SRX043656/SRR197985_1.fastq.bz2
  ftp://ftp.ddbj.nig.ac.jp/ddbj_database/dra/fastq/SRA036/SRA036382/SRX043656/SRR197985_2.fastq.bz2
  ftp://ftp.ddbj.nig.ac.jp/ddbj_database/dra/fastq/SRA036/SRA036383/SRX043656/SRR197986.fastq.bz2
  ftp://ftp.ddbj.nig.ac.jp/ddbj_database/dra/fastq/SRA036/SRA036383/SRX043656/SRR197986_1.fastq.bz2
  ftp://ftp.ddbj.nig.ac.jp/ddbj_database/dra/fastq/SRA036/SRA036383/SRX043656/SRR197986_2.fastq.bz2

- *M. balbisiana*
  ftp://ftp.ddbj.nig.ac.jp/ddbj_database/dra/fastq/SRA098/SRA098922/SRX339427/SRR956987.fastq.bz2

```
ftp://ftp.ddbj.nig.ac.jp/ddbj_database/dra/fastq/SRA098/SRA098922/SRX339427/SRR957627.fastq.bz2
```

- *H. sapiens 1*
  ```
  https://dnanexus-rnd.s3.amazonaws.com/NA12878-xten/reads/NA12878D_HiSeqX_R1.fastq.gz
  https://dnanexus-rnd.s3.amazonaws.com/NA12878-xten/reads/NA12878D_HiSeqX_R2.fastq.gz
  ```
- *H. sapiens 2*
  ```
  ftp.sra.ebi.ac.uk/vol1/fastq/ERR024/ERR024163/ERR024163_1.fastq.gz
  ftp.sra.ebi.ac.uk/vol1/fastq/ERR024/ERR024163/ERR024163_2.fastq.gz
  ftp.sra.ebi.ac.uk/vol1/fastq/ERR024/ERR024164/ERR024164_1.fastq.gz
  ftp.sra.ebi.ac.uk/vol1/fastq/ERR024/ERR024164/ERR024164_2.fastq.gz
  ftp.sra.ebi.ac.uk/vol1/fastq/ERR024/ERR024165/ERR024165_1.fastq.gz
  ftp.sra.ebi.ac.uk/vol1/fastq/ERR024/ERR024165/ERR024165_2.fastq.gz
  ftp.sra.ebi.ac.uk/vol1/fastq/ERR024/ERR024166/ERR024166_1.fastq.gz
  ftp.sra.ebi.ac.uk/vol1/fastq/ERR024/ERR024166/ERR024166_2.fastq.gz
  ftp.sra.ebi.ac.uk/vol1/fastq/ERR024/ERR024167/ERR024167_1.fastq.gz
  ftp.sra.ebi.ac.uk/vol1/fastq/ERR024/ERR024167/ERR024167_2.fastq.gz
  ftp.sra.ebi.ac.uk/vol1/fastq/ERR024/ERR024168/ERR024168_1.fastq.gz
  ftp.sra.ebi.ac.uk/vol1/fastq/ERR024/ERR024168/ERR024168_2.fastq.gz
  ftp.sra.ebi.ac.uk/vol1/fastq/ERR024/ERR024169/ERR024169_1.fastq.gz
  ftp.sra.ebi.ac.uk/vol1/fastq/ERR024/ERR024169/ERR024169_2.fastq.gz
  ftp.sra.ebi.ac.uk/vol1/fastq/ERR024/ERR024170/ERR024170_1.fastq.gz
  ftp.sra.ebi.ac.uk/vol1/fastq/ERR024/ERR024170/ERR024170_2.fastq.gz
  ftp.sra.ebi.ac.uk/vol1/fastq/ERR024/ERR024171/ERR024171_1.fastq.gz
  ftp.sra.ebi.ac.uk/vol1/fastq/ERR024/ERR024171/ERR024171_2.fastq.gz
  ftp.sra.ebi.ac.uk/vol1/fastq/ERR024/ERR024172/ERR024172_1.fastq.gz
  ftp.sra.ebi.ac.uk/vol1/fastq/ERR024/ERR024172/ERR024172_2.fastq.gz
  ftp.sra.ebi.ac.uk/vol1/fastq/ERR024/ERR024173/ERR024173_1.fastq.gz
  ftp.sra.ebi.ac.uk/vol1/fastq/ERR024/ERR024173/ERR024173_2.fastq.gz
  ftp.sra.ebi.ac.uk/vol1/fastq/ERR024/ERR024174/ERR024174_1.fastq.gz
  ftp.sra.ebi.ac.uk/vol1/fastq/ERR024/ERR024174/ERR024174_2.fastq.gz
  ftp.sra.ebi.ac.uk/vol1/fastq/ERR024/ERR024175/ERR024175_1.fastq.gz
  ftp.sra.ebi.ac.uk/vol1/fastq/ERR024/ERR024175/ERR024175_2.fastq.gz
  ftp.sra.ebi.ac.uk/vol1/fastq/ERR024/ERR024176/ERR024176_1.fastq.gz
  ftp.sra.ebi.ac.uk/vol1/fastq/ERR024/ERR024176/ERR024176_2.fastq.gz
  ftp.sra.ebi.ac.uk/vol1/fastq/ERR024/ERR024177/ERR024177_1.fastq.gz
  ftp.sra.ebi.ac.uk/vol1/fastq/ERR024/ERR024177/ERR024177_2.fastq.gz
  ftp.sra.ebi.ac.uk/vol1/fastq/ERR024/ERR024178/ERR024178_1.fastq.gz
  ftp.sra.ebi.ac.uk/vol1/fastq/ERR024/ERR024178/ERR024178_2.fastq.gz
  ftp.sra.ebi.ac.uk/vol1/fastq/ERR024/ERR024179/ERR024179_1.fastq.gz
  ftp.sra.ebi.ac.uk/vol1/fastq/ERR024/ERR024179/ERR024179_2.fastq.gz
  ftp.sra.ebi.ac.uk/vol1/fastq/ERR024/ERR024180/ERR024180_1.fastq.gz
  ftp.sra.ebi.ac.uk/vol1/fastq/ERR024/ERR024180/ERR024180_2.fastq.gz
  ftp.sra.ebi.ac.uk/vol1/fastq/ERR024/ERR024181/ERR024181_1.fastq.gz
  ftp.sra.ebi.ac.uk/vol1/fastq/ERR024/ERR024181/ERR024181_2.fastq.gz
  ftp.sra.ebi.ac.uk/vol1/fastq/ERR024/ERR024182/ERR024182_1.fastq.gz
  ftp.sra.ebi.ac.uk/vol1/fastq/ERR024/ERR024182/ERR024182_2.fastq.gz
  ftp.sra.ebi.ac.uk/vol1/fastq/ERR024/ERR024183/ERR024183_1.fastq.gz
  ftp.sra.ebi.ac.uk/vol1/fastq/ERR024/ERR024183/ERR024183_2.fastq.gz
  ftp.sra.ebi.ac.uk/vol1/fastq/ERR024/ERR024184/ERR024184_1.fastq.gz
  ftp.sra.ebi.ac.uk/vol1/fastq/ERR024/ERR024184/ERR024184_2.fastq.gz
  ftp.sra.ebi.ac.uk/vol1/fastq/ERR024/ERR024185/ERR024185_1.fastq.gz
  ftp.sra.ebi.ac.uk/vol1/fastq/ERR024/ERR024185/ERR024185_2.fastq.gz
  ftp.sra.ebi.ac.uk/vol1/fastq/ERR024/ERR024186/ERR024186_1.fastq.gz
  ```

ftp.sra.ebi.ac.uk/vol1/fastq/ERR024/ERR024186/ERR024186_2.fastq.gz

- *H. sapiens 3*
  ftp.sra.ebi.ac.uk/vol1/fastq/ERR174/ERR174324/ERR174324_1.fastq.gz
  ftp.sra.ebi.ac.uk/vol1/fastq/ERR174/ERR174324/ERR174324_2.fastq.gz
  ftp.sra.ebi.ac.uk/vol1/fastq/ERR174/ERR174325/ERR174325_1.fastq.gz
  ftp.sra.ebi.ac.uk/vol1/fastq/ERR174/ERR174325/ERR174325_2.fastq.gz
  ftp.sra.ebi.ac.uk/vol1/fastq/ERR174/ERR174326/ERR174326_1.fastq.gz
  ftp.sra.ebi.ac.uk/vol1/fastq/ERR174/ERR174326/ERR174326_2.fastq.gz
  ftp.sra.ebi.ac.uk/vol1/fastq/ERR174/ERR174327/ERR174327_1.fastq.gz
  ftp.sra.ebi.ac.uk/vol1/fastq/ERR174/ERR174327/ERR174327_2.fastq.gz
  ftp.sra.ebi.ac.uk/vol1/fastq/ERR174/ERR174328/ERR174328_1.fastq.gz
  ftp.sra.ebi.ac.uk/vol1/fastq/ERR174/ERR174328/ERR174328_2.fastq.gz
  ftp.sra.ebi.ac.uk/vol1/fastq/ERR174/ERR174329/ERR174329_1.fastq.gz
  ftp.sra.ebi.ac.uk/vol1/fastq/ERR174/ERR174329/ERR174329_2.fastq.gz
  ftp.sra.ebi.ac.uk/vol1/fastq/ERR174/ERR174330/ERR174330_1.fastq.gz
  ftp.sra.ebi.ac.uk/vol1/fastq/ERR174/ERR174330/ERR174330_2.fastq.gz
  ftp.sra.ebi.ac.uk/vol1/fastq/ERR174/ERR174331/ERR174331_1.fastq.gz
  ftp.sra.ebi.ac.uk/vol1/fastq/ERR174/ERR174331/ERR174331_2.fastq.gz
  ftp.sra.ebi.ac.uk/vol1/fastq/ERR174/ERR174332/ERR174332_1.fastq.gz
  ftp.sra.ebi.ac.uk/vol1/fastq/ERR174/ERR174332/ERR174332_2.fastq.gz
  ftp.sra.ebi.ac.uk/vol1/fastq/ERR174/ERR174333/ERR174333_1.fastq.gz
  ftp.sra.ebi.ac.uk/vol1/fastq/ERR174/ERR174333/ERR174333_2.fastq.gz
  ftp.sra.ebi.ac.uk/vol1/fastq/ERR174/ERR174334/ERR174334_1.fastq.gz
  ftp.sra.ebi.ac.uk/vol1/fastq/ERR174/ERR174334/ERR174334_2.fastq.gz
  ftp.sra.ebi.ac.uk/vol1/fastq/ERR174/ERR174335/ERR174335_1.fastq.gz
  ftp.sra.ebi.ac.uk/vol1/fastq/ERR174/ERR174335/ERR174335_2.fastq.gz
  ftp.sra.ebi.ac.uk/vol1/fastq/ERR174/ERR174336/ERR174336_1.fastq.gz
  ftp.sra.ebi.ac.uk/vol1/fastq/ERR174/ERR174336/ERR174336_2.fastq.gz
  ftp.sra.ebi.ac.uk/vol1/fastq/ERR174/ERR174337/ERR174337_1.fastq.gz
  ftp.sra.ebi.ac.uk/vol1/fastq/ERR174/ERR174337/ERR174337_2.fastq.gz
  ftp.sra.ebi.ac.uk/vol1/fastq/ERR174/ERR174338/ERR174338_1.fastq.gz
  ftp.sra.ebi.ac.uk/vol1/fastq/ERR174/ERR174338/ERR174338_2.fastq.gz
  ftp.sra.ebi.ac.uk/vol1/fastq/ERR174/ERR174339/ERR174339_1.fastq.gz
  ftp.sra.ebi.ac.uk/vol1/fastq/ERR174/ERR174339/ERR174339_2.fastq.gz
  ftp.sra.ebi.ac.uk/vol1/fastq/ERR174/ERR174340/ERR174340_1.fastq.gz
  ftp.sra.ebi.ac.uk/vol1/fastq/ERR174/ERR174340/ERR174340_2.fastq.gz
  ftp.sra.ebi.ac.uk/vol1/fastq/ERR174/ERR174341/ERR174341_1.fastq.gz
  ftp.sra.ebi.ac.uk/vol1/fastq/ERR174/ERR174341/ERR174341_2.fastq.gz

## 3.2 $k$-mer counting for various datasets

In this section the detailed results of the examined $k$-mer counting tools are given.

Sequencing errors usually result in single-base substitution/insertion/deletion in reads. Each of such errors generates a number (up to $k$) of $k$-mers that likely appear only once in the $k$-mer database. Thus, such rare $k$-mers are usually filtered out during the $k$-mer counting. This subsection involves results of programs run with enabled single-appearance $k$-mers removal. Resignation from this filtering has small impact on the running time (and memory usage) of the examined algorithms, usually less than 10 percent.

All modern workstations and servers are equipped with one (or more) multicore CPU. Thus, to compare the tools in a real environment the experiments were performed using several threads: 12 (the no. of cores in a single CPU) for Xeon-based workstation and 16 for Opteron-based server. The only exception is the KCMBT algorithm that requires the number of threads to be a power of 2. Thus, KCMBT was executed using 8 threads at Xeon-based workstation. We decided not to use 16 threads here to reduce the huge memory requirements of KCMBT (exceeding amount of RAM at Xeon-based workstation for most of test datasets when 16 threads are used).

The FASTQ files are huge, so they are almost always stored in a compressed form (usually they are gzipped). Thus, many tools allow to directly process gzipped FASTQ files. The other require prior decompression (spending time and space). We present the running times for both uncompressed and compressed input files, but clearly the more important are the ones for the gzipped FASTQ ('gz-Time' column).

Table 2: Comparison of $k$-mer counting algorithms for *F. vesca* dataset. The times are given for: uncompressed input FASTQ files ('Time') and gzipped input FASTQ files ('gz-Time'). The units are: seconds (time), GB (Disk and RAM). All programs were executed in 12-threads mode (Xeon) / 16-threads mode (Opteron). '—' means that the mode is not supported.

| Algorithm | $k=28$ | | | $k=40$ | | | $k=55$ | | | $k=65$ | | |
|---|---|---|---|---|---|---|---|---|---|---|---|---|
| | RAM | Disk | Time/gz-Time | RAM | Disk | Time/gz-Time | RAM | Disk | Time/gz-Time | RAM | Disk | Time/gz-Time |
| **Xeon HDD** | | | | | | | | | | | | |
| DSK 2 | 5 | 6 | 78 / 147 | 7 | 7 | 99 / 181 | 7 | 5 | 101 / 180 | 9 | 6 | 151 / 209 |
| Gerbil | 1 | 4 | 52 / 58 | 1 | 3 | 86 / 74 | 1 | 3 | 105 / 90 | 2 | 3 | 80 / 78 |
| GTester4 | 28 | 0 | 328 / — | \multicolumn{3}{c}{unsupported $k$} | | | | \multicolumn{3}{c}{unsupported $k$} | | |
| Jellyfish 2 | 8 | 0 | 147 / — | 14 | 0 | 203 / — | 39 | 0 | 325 / — | 47 | 0 | 350 / — |
| KCMBT | 28 | 0 | 146 / — | \multicolumn{3}{c}{unsupported $k$} | | | | \multicolumn{3}{c}{unsupported $k$} | | |
| KMC 2 | 7 | 4 | 51 / 44 | 12 | 3 | 91 / 80 | 12 | 3 | 102 / 93 | 12 | 3 | 141 / 130 |
| KMC 3 | 9 | 4 | 48 / 40 | 12 | 4 | 60 / 44 | 12 | 3 | 61 / 43 | 12 | 3 | 69 / 49 |
| **Xeon SSD** | | | | | | | | | | | | |
| DSK 2 | 5 | 6 | 70 / 146 | 6 | 7 | 98 / 181 | 7 | 5 | 99 / 180 | 9 | 6 | 144 / 209 |
| Gerbil | 1 | 4 | 43 / 67 | 1 | 3 | 69 / 84 | 1 | 3 | 98 / 106 | 2 | 3 | 74 / 88 |
| GTester4 | 29 | 0 | 319 / — | \multicolumn{3}{c}{unsupported $k$} | | | | \multicolumn{3}{c}{unsupported $k$} | | |
| Jellyfish 2 | 9 | 0 | 143 / — | 13 | 0 | 188 / — | 39 | 0 | 264 / — | 47 | 0 | 277 / — |
| KCMBT | 27 | 0 | 63 / — | \multicolumn{3}{c}{unsupported $k$} | | | | \multicolumn{3}{c}{unsupported $k$} | | |
| KMC 2 | 7 | 4 | 42 / 58 | 12 | 3 | 89 / 118 | 12 | 3 | 107 / 143 | 12 | 3 | 131 / 220 |
| KMC 3 | 9 | 4 | 44 / 45 | 12 | 4 | 58 / 56 | 12 | 3 | 56 / 54 | 12 | 3 | 75 / 70 |
| **Opteron HDD** | | | | | | | | | | | | |
| DSK 2 | 7 | 6 | 83 / 160 | 11 | 7 | 117 / 196 | 10 | 5 | 123 / 195 | 10 | 6 | 151 / 229 |
| Gerbil | 1 | 4 | 53 / 95 | 1 | 3 | 87 / 127 | 1 | 3 | 121 / 129 | 2 | 3 | 107 / 121 |
| GTester4 | 29 | 0 | 323 / — | \multicolumn{3}{c}{unsupported $k$} | | | | \multicolumn{3}{c}{unsupported $k$} | | |
| Jellyfish 2 | 9 | 0 | 227 / — | 13 | 0 | 350 / — | 39 | 0 | 331 / — | 47 | 0 | 404 / — |
| KCMBT | 22 | 0 | 191 / — | \multicolumn{3}{c}{unsupported $k$} | | | | \multicolumn{3}{c}{unsupported $k$} | | |
| KMC 2 | 9 | 4 | 61 / 61 | 12 | 3 | 106 / 108 | 12 | 3 | 123 / 128 | 12 | 3 | 202 / 196 |
| KMC 3 | 11 | 4 | 55 / 56 | 12 | 4 | 67 / 61 | 12 | 3 | 67 / 60 | 12 | 3 | 86 / 77 |

15

Table 3: Comparison of $k$-mer counting algorithms for *G. gallus* dataset. The times are given for: uncompressed input FASTQ files ('Time') and gzipped input FASTQ files ('gz-Time'). The units are: seconds (time), GB (Disk and RAM). All programs were executed in 12-threads mode (Xeon) / 16-threads mode (Opteron). '—' means that the mode is not supported.

| Algorithm | $k = 28$ | | | $k = 40$ | | | $k = 55$ | | | $k = 65$ | | |
|---|---|---|---|---|---|---|---|---|---|---|---|---|
| | RAM | Disk | Time/gz-Time | RAM | Disk | Time/gz-Time | RAM | Disk | Time/gz-Time | RAM | Disk | Time/gz-Time |
| **Xeon HDD** | | | | | | | | | | | | |
| DSK 2 | 14 | 39 | 783 / 1,180 | 15 | 48 | 894 / 1,257 | 15 | 30 | 719 / 1,123 | 14 | 35 | 845 / 1,188 |
| Gerbil | 2 | 24 | 607 / 631 | 5 | 21 | 664 / 585 | 4 | 17 | 615 / 551 | 5 | 15 | 575 / 501 |
| GTester4 | 74 | 0 | 2,494 / — | *unsupported k* | | | *unsupported k* | | | *unsupported k* | | |
| Jellyfish 2 | 33 | 0 | 909 / — | 52 | 0 | 1,099 / — | 77 | 0 | 1,048 / — | 93 | 0 | 938 / — |
| KCMBT | 107 | 0 | 1,335 / — | *unsupported k* | | | *unsupported k* | | | *unsupported k* | | |
| KMC 2 | 12 | 25 | 489 / 321 | 12 | 21 | 792 / 606 | 12 | 18 | 861 / 672 | 12 | 16 | 1,099 / 850 |
| KMC 3 | 12 | 28 | 492 / 292 | 12 | 22 | 496 / 268 | 12 | 18 | 455 / 245 | 12 | 16 | 462 / 253 |
| **Xeon SSD** | | | | | | | | | | | | |
| DSK 2 | 13 | 39 | 663 / 1,107 | 15 | 48 | 775 / 1,142 | 14 | 30 | 688 / 1,055 | 14 | 35 | 809 / 1,112 |
| Gerbil | 2 | 24 | 452 / 445 | 5 | 21 | 512 / 492 | 4 | 17 | 487 / 491 | 5 | 15 | 451 / 424 |
| GTester4 | 132 | 0 | 2,429 / — | *unsupported k* | | | *unsupported k* | | | *unsupported k* | | |
| Jellyfish 2 | 33 | 0 | 836 / — | 52 | 0 | 1,013 / — | 77 | 0 | 963 / — | 93 | 0 | 805 / — |
| KCMBT | 108 | 0 | 785 / — | *unsupported k* | | | *unsupported k* | | | *unsupported k* | | |
| KMC 2 | 12 | 25 | 386 / 251 | 12 | 21 | 777 / 559 | 12 | 18 | 837 / 610 | 12 | 16 | 1,021 / 780 |
| KMC 3 | 12 | 28 | 370 / 250 | 12 | 22 | 394 / 259 | 12 | 18 | 359 / 231 | 12 | 16 | 367 / 249 |
| **Opteron HDD** | | | | | | | | | | | | |
| DSK 2 | 15 | 39 | 860 / 1,528 | 15 | 48 | 1,014 / 1,598 | 14 | 30 | 932 / 1,504 | 14 | 35 | 982 / 1,519 |
| Gerbil | 3 | 24 | 655 / 824 | 5 | 21 | 785 / 850 | 4 | 17 | 672 / 795 | 5 | 15 | 577 / 735 |
| GTester4 | 162 | 0 | 2,500 / — | *unsupported k* | | | *unsupported k* | | | *unsupported k* | | |
| Jellyfish 2 | 33 | 0 | 1,439 / — | 52 | 0 | 1,335 / — | 77 | 0 | 1,124 / — | 93 | 0 | 1,124 / — |
| KCMBT | 168 | 0 | 1,606 / — | *unsupported k* | | | *unsupported k* | | | *unsupported k* | | |
| KMC 2 | 12 | 25 | 509 / 371 | 12 | 21 | 702 / 641 | 12 | 18 | 764 / 668 | 12 | 16 | 937 / 840 |
| KMC 3 | 12 | 28 | 470 / 272 | 12 | 22 | 504 / 320 | 12 | 18 | 471 / 277 | 12 | 16 | 514 / 323 |

16

Table 4: Comparison of $k$-mer counting algorithms for *M. balbisiana* dataset. The times are given for: uncompressed input FASTQ files ('Time') and gzipped input FASTQ files ('gz-Time'). The units are: seconds (time), GB (Disk and RAM). All programs were executed in 12-threads mode (Xeon) / 16-threads mode (Opteron). '—' means that the mode is not supported.

| Algorithm | $k=28$ RAM | Disk | Time/gz-Time | $k=40$ RAM | Disk | Time/gz-Time | $k=55$ RAM | Disk | Time/gz-Time | $k=65$ RAM | Disk | Time/gz-Time |
|---|---|---|---|---|---|---|---|---|---|---|---|---|
| **Xeon HDD** | | | | | | | | | | | | |
| DSK 2 | 14 | 64 | 1,350 / 1,851 | 13 | 76 | 1,397 / 1,995 | 12 | 49 | 1,108 / 1,744 | 12 | 57 | 1,264 / 1,847 |
| Gerbil | 2 | 39 | 943 / 918 | 4 | 33 | 940 / 896 | 4 | 27 | 885 / 842 | 5 | 24 | 849 / 814 |
| GTester4 | 119 | 0 | 3,645 / — | unsupported $k$ | | | unsupported $k$ | | | unsupported $k$ | | |
| Jellyfish 2 | 17 | 0 | 1,152 / — | 17 | 0 | 1,467 / — | 26 | 0 | 1,281 / — | 31 | 0 | 662 / — |
| KCMBT | 67 | 0 | 2,058 / — | unsupported $k$ | | | unsupported $k$ | | | unsupported $k$ | | |
| KMC 2 | 12 | 41 | 857 / 845 | 12 | 35 | 1,040 / 1,033 | 12 | 29 | 1,046 / 1,046 | 12 | 25 | 1,123 / 1,138 |
| KMC 3 | 12 | 45 | 820 / 671 | 12 | 36 | 808 / 681 | 12 | 29 | 771 / 658 | 12 | 26 | 773 / 668 |
| **Xeon SSD** | | | | | | | | | | | | |
| DSK 2 | 14 | 64 | 983 / 1,705 | 13 | 76 | 1,092 / 1,754 | 12 | 49 | 959 / 1,671 | 12 | 57 | 1,106 / 1,756 |
| Gerbil | 2 | 39 | 713 / 718 | 4 | 33 | 692 / 727 | 4 | 27 | 650 / 677 | 5 | 24 | 615 / 650 |
| GTester4 | 132 | 0 | 3,768 / — | unsupported $k$ | | | unsupported $k$ | | | unsupported $k$ | | |
| Jellyfish 2 | 17 | 0 | 1,120 / — | 17 | 0 | 1,394 / — | 26 | 0 | 1,315 / — | 31 | 0 | 643 / — |
| KCMBT | 68 | 0 | 486 / — | unsupported $k$ | | | unsupported $k$ | | | unsupported $k$ | | |
| KMC 2 | 12 | 41 | 664 / 715 | 12 | 35 | 861 / 939 | 12 | 29 | 847 / 949 | 12 | 25 | 935 / 1,017 |
| KMC 3 | 12 | 45 | 622 / 653 | 12 | 36 | 639 / 670 | 12 | 29 | 603 / 647 | 12 | 26 | 605 / 660 |
| **Opteron HDD** | | | | | | | | | | | | |
| DSK 2 | 13 | 64 | 1,350 / 2,291 | 13 | 76 | 1,708 / 2,423 | 13 | 49 | 1,397 / 2,224 | 14 | 57 | 1,513 / 2,229 |
| Gerbil | 2 | 39 | 962 / 1,223 | 4 | 33 | 1,027 / 1,197 | 4 | 27 | 906 / 1,025 | 5 | 24 | 904 / 1,006 |
| GTester4 | 221 | 0 | 3,704 / — | unsupported $k$ | | | unsupported $k$ | | | unsupported $k$ | | |
| Jellyfish 2 | 17 | 0 | 2,086 / — | 17 | 0 | 3,846 / — | 26 | 0 | 1,184 / — | 31 | 0 | 1,087 / — |
| KCMBT | 122 | 0 | 2,573 / — | unsupported $k$ | | | unsupported $k$ | | | unsupported $k$ | | |
| KMC 2 | 12 | 41 | 849 / 1,007 | 12 | 35 | 1,229 / 1,396 | 12 | 29 | 1,327 / 1,376 | 12 | 25 | 1,736 / 1,672 |
| KMC 3 | 12 | 45 | 745 / 709 | 12 | 36 | 823 / 820 | 12 | 29 | 754 / 766 | 12 | 26 | 833 / 828 |

17

Table 5: Comparison of k-mer counting algorithms for *H. sapiens* 1 dataset. The times are given for: uncompressed input FASTQ files ('Time') and gzipped input FASTQ files ('gz-Time'). The units are: seconds (time), GB (Disk and RAM). All programs were executed in 12-threads mode (Xeon) / 16-threads mode (Opteron). '—' means that the mode is not supported.

| Algorithm | k = 28 | | | | k = 40 | | | | k = 55 | | | | k = 65 | | | |
|---|---|---|---|---|---|---|---|---|---|---|---|---|---|---|---|---|
| | RAM | Disk | Time/gz-Time | | RAM | Disk | Time/gz-Time | | RAM | Disk | Time/gz-Time | | RAM | Disk | Time/gz-Time | |
| **Xeon HDD** | | | | | | | | | | | | | | | | |
| DSK 2 | 13 | 144 | 3,237 / 4,036 | | 14 | 174 | 3,938 / 5,097 | | 15 | 112 | 3,143 / 4,213 | | 15 | 134 | 4,325 / 5,616 | |
| Gerbil | 7 | 89 | 1,883 / 1,876 | | 15 | 76 | 2,047 / 2,069 | | 17 | 66 | 1,901 / 1,957 | | 24 | 61 | 1,844 / 1,915 | |
| GTester4 | Out of memory | | | | unsupported k | | | | unsupported k | | | | unsupported k | | | |
| Jellyfish 2 | 125 | 0 | 4,769 / — | | 103 | 261 | 9,206 / — | | 52 | 548 | 14,176 / — | | 62 | 617 | 14,597 / — | |
| KCMBT | Out of memory | | | | unsupported k | | | | unsupported k | | | | unsupported k | | | |
| KMC 2 | 12 | 94 | 1,774 / 1,829 | | 13 | 80 | 3,127 / 3,138 | | 15 | 69 | 3,370 / 3,387 | | 22 | 64 | 4,351 / 4,378 | |
| KMC 3 | 12 | 101 | 1,507 / 1,194 | | 12 | 83 | 1,542 / 1,264 | | 12 | 71 | 1,472 / 1,217 | | 12 | 65 | 1,574 / 1,331 | |
| **Xeon SSD** | | | | | | | | | | | | | | | | |
| DSK 2 | 13 | 144 | 2,091 / 3,340 | | 15 | 174 | 3,114 / 4,316 | | 15 | 112 | 2,797 / 3,956 | | 16 | 134 | 3,631 / 5,440 | |
| Gerbil | 7 | 89 | 1,543 / 1,710 | | 15 | 76 | 1,765 / 1,959 | | 17 | 66 | 1,636 / 1,915 | | 25 | 61 | 1,520 / 1,843 | |
| GTester4 | Out of memory | | | | unsupported k | | | | unsupported k | | | | unsupported k | | | |
| Jellyfish 2 | 124 | 0 | 4,640 / — | | 106 | 261 | 14,969 / — | | Out of disk | | | | Out of disk | | | |
| KCMBT | Out of memory | | | | unsupported k | | | | unsupported k | | | | unsupported k | | | |
| KMC 2 | 12 | 94 | 1,494 / 1,593 | | 13 | 80 | 2,861 / 2,932 | | 15 | 69 | 2,964 / 3,231 | | 22 | 64 | 4,080 / 4,248 | |
| KMC 3 | 12 | 101 | 1,184 / 1,158 | | 12 | 83 | 1,207 / 1,254 | | 12 | 71 | 1,157 / 1,203 | | 12 | 65 | 1,241 / 1,316 | |
| **Opteron HDD** | | | | | | | | | | | | | | | | |
| DSK 2 | 12 | 144 | 4,326 / 4,725 | | 15 | 174 | 4,594 / 6,331 | | 15 | 112 | 3,750 / 5,146 | | 15 | 134 | 4,467 / 6,067 | |
| Gerbil | 7 | 89 | 1,893 / 3,107 | | 15 | 76 | 2,132 / 3,074 | | 17 | 66 | 2,072 / 2,650 | | 25 | 61 | 2,054 / 2,479 | |
| GTester4 | 268 | 0 | 13,125 / — | | unsupported k | | | | unsupported k | | | | unsupported k | | | |
| Jellyfish 2 | 123 | 0 | 5,164 / — | | 204 | 0 | 6,642 / — | | 153 | 347 | 10,939 / — | | 185 | 399 | 11,247 / — | |
| KCMBT | Out of memory | | | | unsupported k | | | | unsupported k | | | | unsupported k | | | |
| KMC 2 | 12 | 94 | 1,786 / 1,997 | | 13 | 80 | 3,236 / 3,344 | | 15 | 69 | 3,346 / 3,711 | | 22 | 64 | 5,302 / 5,550 | |
| KMC 3 | 12 | 101 | 1,328 / 1,276 | | 12 | 83 | 1,607 / 1,610 | | 12 | 71 | 1,520 / 1,468 | | 12 | 65 | 1,821 / 1,824 | |

18

Table 6: Comparison of $k$-mer counting algorithms for *H. sapiens* 2 dataset. The times are given for: uncompressed input FASTQ files ('Time') and gzipped input FASTQ files ('gz-Time'). The units are: seconds (time), GB (Disk and RAM). All programs were executed in 12-threads mode (Xeon) / 16-threads mode (Opteron). '—' means that the mode is not supported.

| Algorithm | $k=28$ | | | $k=40$ | | | $k=55$ | | | $k=65$ | | |
|---|---|---|---|---|---|---|---|---|---|---|---|---|
| | RAM | Disk | Time/gz-Time | RAM | Disk | Time/gz-Time | RAM | Disk | Time/gz-Time | RAM | Disk | Time/gz-Time |
| **Xeon HDD** | | | | | | | | | | | | |
| DSK 2 | 14 | 157 | 3,543 / 4,965 | 15 | 185 | 4,085 / 5,650 | 15 | 118 | 3,257 / 4,889 | 14 | 138 | 3,801 / 5,392 |
| Gerbil | 6 | 96 | 2,165 / 2,177 | 12 | 80 | 2,212 / 2,313 | 12 | 67 | 2,038 / 2,114 | 17 | 59 | 1,872 / 1,950 |
| GTester4 | 114 | 0 | 14,253 / — | *unsupported k* | | | *unsupported k* | | | *unsupported k* | | |
| Jellyfish 2 | 63 | 0 | 3,337 / — | 102 | 104 | 9,550 / — | 52 | 327 | 9,504 / — | 62 | 345 | 8,702 / — |
| KCMBT | *Out of memory* | | | *unsupported k* | | | *unsupported k* | | | *unsupported k* | | |
| KMC 2 | 12 | 101 | 1,998 / 1,571 | 13 | 84 | 3,032 / 2,704 | 14 | 70 | 3,074 / 2,764 | 18 | 62 | 3,478 / 3,148 |
| KMC 3 | 12 | 109 | 1,800 / 1,175 | 12 | 87 | 1,788 / 1,139 | 12 | 71 | 1,714 / 1,011 | 12 | 63 | 1,704 / 1,013 |
| **Xeon SSD** | | | | | | | | | | | | |
| DSK 2 | 14 | 157 | 2,273 / 4,388 | 15 | 185 | 3,155 / 4,848 | 15 | 118 | 2,764 / 4,444 | 14 | 138 | 3,136 / 4,833 |
| Gerbil | 6 | 96 | 1,692 / 1,928 | 12 | 80 | 1,676 / 2,039 | 13 | 67 | 1,529 / 1,904 | 17 | 59 | 1,360 / 1,769 |
| GTester4 | 132 | 0 | 12,301 / — | *unsupported k* | | | *unsupported k* | | | *unsupported k* | | |
| Jellyfish 2 | 63 | 0 | 3,285 / — | 102 | 104 | 10,776 / — | 52 | 327 | 12,698 / — | 62 | 345 | 11,571 / — |
| KCMBT | *Out of memory* | | | *unsupported k* | | | *unsupported k* | | | *unsupported k* | | |
| KMC 2 | 12 | 101 | 1,583 / 1,165 | 13 | 84 | 2,756 / 2,266 | 14 | 70 | 2,654 / 2,397 | 18 | 62 | 3,115 / 2,785 |
| KMC 3 | 12 | 109 | 1,289 / 963 | 12 | 87 | 1,334 / 957 | 12 | 71 | 1,244 / 821 | 12 | 63 | 1,227 / 844 |
| **Opteron HDD** | | | | | | | | | | | | |
| DSK 2 | 15 | 157 | 5,475 / 7,030 | 15 | 185 | 6,110 / 8,067 | 15 | 118 | 4,760 / 6,261 | 15 | 138 | 4,307 / 6,759 |
| Gerbil | 6 | 96 | 2,203 / 3,078 | 12 | 80 | 2,143 / 3,202 | 13 | 67 | 2,133 / 2,597 | 17 | 59 | 1,944 / 2,322 |
| GTester4 | 269 | 0 | 11,891 / — | *unsupported k* | | | *unsupported k* | | | *unsupported k* | | |
| Jellyfish 2 | 63 | 0 | 5,496 / — | 204 | 0 | 5,595 / — | 153 | 0 | 7,194 / — | 185 | 0 | 4,071 / — |
| KCMBT | *Out of memory* | | | *unsupported k* | | | *unsupported k* | | | *unsupported k* | | |
| KMC 2 | 13 | 101 | 2,205 / 1,485 | 13 | 84 | 3,361 / 2,897 | 14 | 70 | 3,233 / 2,743 | 18 | 62 | 3,909 / 3,463 |
| KMC 3 | 12 | 109 | 1,780 / 1,082 | 12 | 87 | 1,950 / 1,311 | 12 | 71 | 1,720 / 1,137 | 12 | 63 | 1,886 / 1,335 |

Table 7: Comparison of $k$-mer counting algorithms for *H. sapiens* 3 dataset. The times are given for: uncompressed input FASTQ files ('Time') and gzipped input FASTQ files ('gz-Time'). The units are: seconds (time), GB (Disk and RAM). All programs were executed in 12-threads mode (Xeon) / 16-threads mode (Opteron). '—' means that the mode is not supported. There are no experiments at Xeon-SSD configuration due to too large input files and limited capacity of the SSD at the test platform.

| Algorithm | $k = 28$ | | | $k = 40$ | | | $k = 55$ | | | $k = 65$ | | |
|---|---|---|---|---|---|---|---|---|---|---|---|---|
| | RAM | Disk | Time/gz-Time | RAM | Disk | Time/gz-Time | RAM | Disk | Time/gz-Time | RAM | Disk | Time/gz-Time |
| **Xeon HDD** | | | | | | | | | | | | |
| DSK 2 | | | *Out of time (>12h)* | | | *Out of time (>12h)* | | | *Out of time (>12h)* | | | *Out of time (>12h)* |
| Gerbil | 29 | 523 | 11,994 / 12,730 | 56 | 436 | 13,328 / 13,725 | 62 | 364 | 11,968 / 12,469 | 72 | 323 | 11,341 / 11,496 |
| GTester4 | | | *Out of time (>12h)* | | | *unsupported k* | | | *unsupported k* | | | *unsupported k* |
| Jellyfish 2 | 121 | 251 | 43,005 / — | | | *Out of time (>12h)* | | | *Out of time (>12h)* | | | *Out of time (>12h)* |
| KCMBT | | | *Out of memory* | | | *unsupported k* | | | *unsupported k* | | | *unsupported k* |
| KMC 2 (64GB) | 64 | 551 | 10,777 / 9,036 | 69 | 457 | 14,732 / 13,041 | 72 | 381 | 13,774 / 11,804 | 96 | 338 | 17,646 / 14,951 |
| KMC 3 (32GB) | 33 | 596 | 9,775 / 5,985 | 36 | 477 | 9,291 / 6,018 | 32 | 389 | 8,481 / 5,331 | 34 | 343 | 8,578 / 5,331 |
| **Opteron HDD** | | | | | | | | | | | | |
| DSK 2 | | | *Out of time (>12h)* | | | *Out of time (>12h)* | | | *Out of time (>12h)* | | | *Out of time (>12h)* |
| Gerbil | 29 | 523 | 11,447 / 18,980 | 56 | 436 | 12,804 / 19,518 | 62 | 364 | 12,559 / 15,549 | 74 | 323 | 11,919 / 13,842 |
| GTester4 | | | *Out of time (>12h)* | | | *unsupported k* | | | *unsupported k* | | | *unsupported k* |
| Jellyfish 2 | 240 | 0 | 32,541 / — | 204 | 369 | 34,618 / — | 153 | 636 | 34,677 / — | 185 | 685 | 31,693 / — |
| KCMBT | | | *Out of memory* | | | *unsupported k* | | | *unsupported k* | | | *unsupported k* |
| KMC 2 (64GB) | 65 | 551 | 11,294 / 9,066 | 69 | 457 | 17,214 / 15,680 | 72 | 381 | 16,701 / 14,654 | 96 | 338 | 22,013 / 20,867 |
| KMC 3 (32GB) | 33 | 596 | 9,558 / 7,046 | 36 | 477 | 10,013 / 7,832 | 32 | 389 | 9,282 / 7,148 | 34 | 343 | 9,557 / 7,961 |

20

## 3.3 Scalability of $k$-mer counting tools

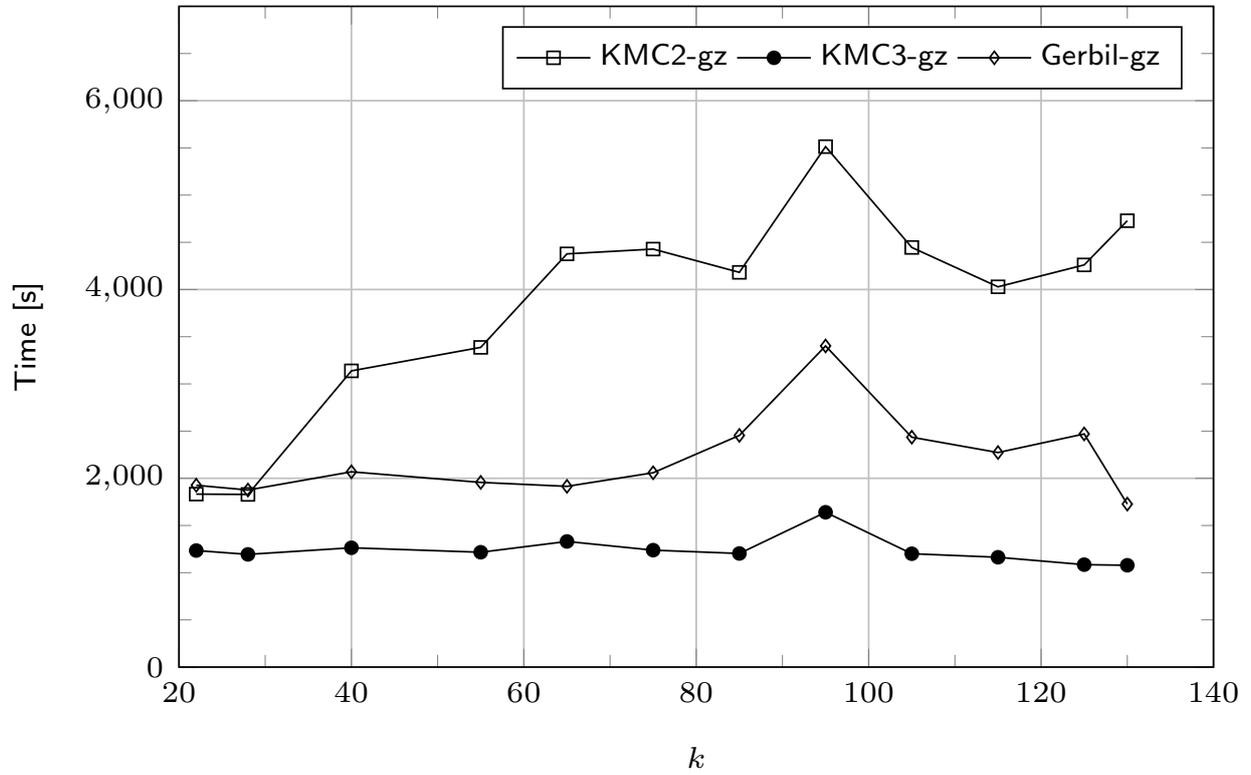

Figure 1: Comparison of running times of the best (fastest) algorithms for $k$-mer counting at Xeon-based workstation (HDD).

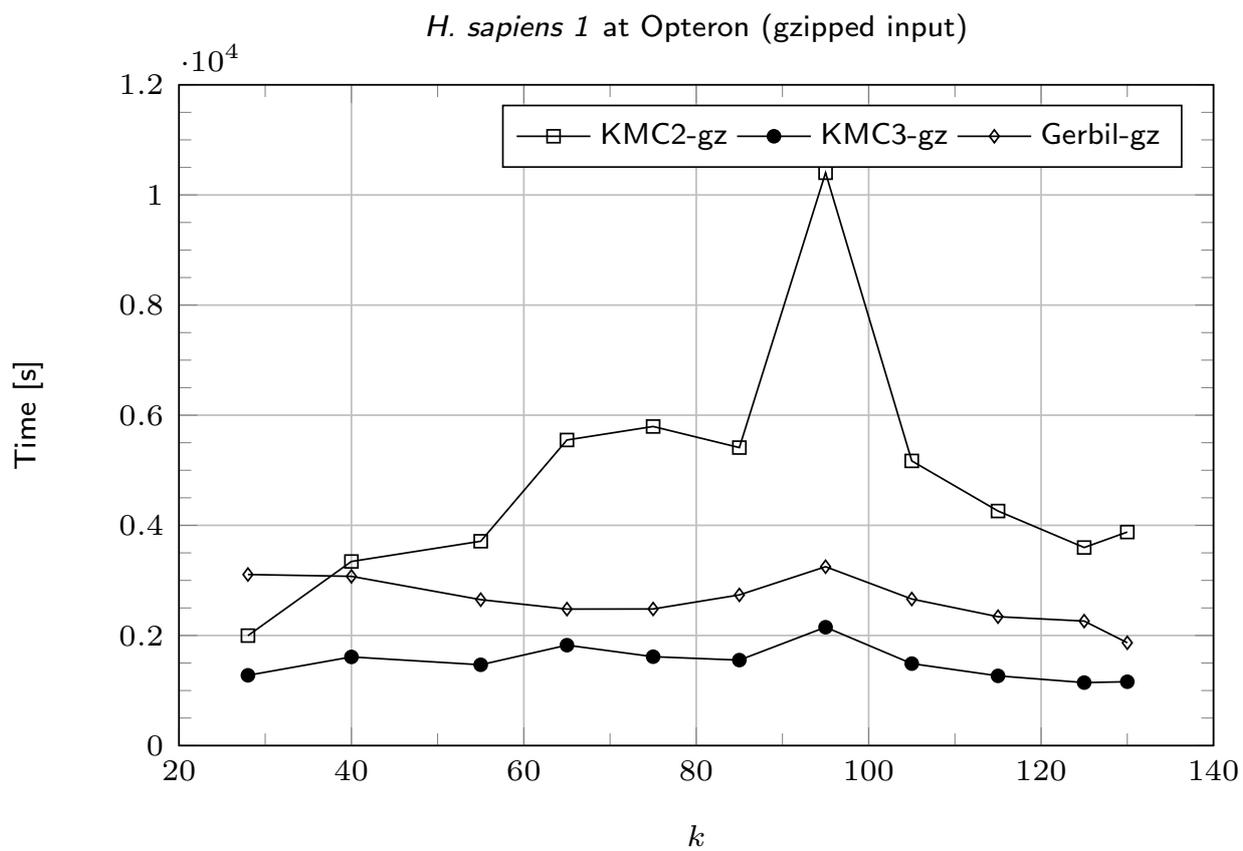

Figure 2: Comparison of running times of the best (fastest) algorithms for $k$-mer counting at Opteron-based server.

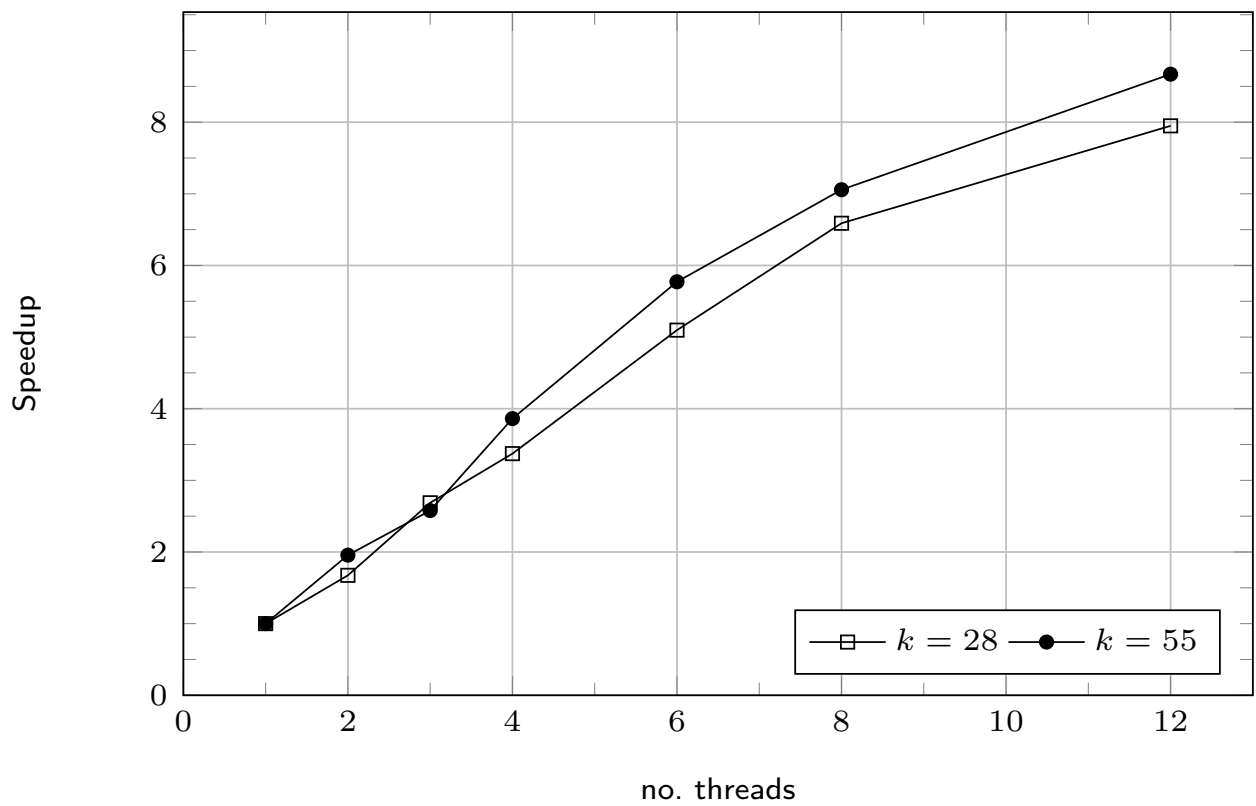

Figure 3: Comparison of scalability of KMC3 on the number of threads employed for processing at Xeon-based workstation (HDD, gzipped input)

# 4 KMC tools

In this section we evaluate KMC tools in practice. To this end we examine how KMC tools can be used to perform some stage of analyzes in three real problems recently discussed in the literature:

- DIAMUND—detection of *de novo* mutations causing genetic diseases in a child of healthy parents [3],
- NIKS—direct comparison (for mutation detection) of two highly related WGS samples [2],
- GenomeTester4—determination of bacteria-group-specific $k$-mers [1].

All experiments in this section were executed at Opteron-based server.

The scripts we used in the experiments described in this section as well as KMC tools documentation can be downloaded from:
http://sun.aei.polsl.pl/REFRESH/index.php?page=projects&project=kmc&subpage=download

## 4.1 DIAMUND

### 4.1.1 Datasets

Table 8: Datasets used in the experiments. No. of bases are in Gbases. File sizes are in Gbytes (1Gbyte = $10^9$ bytes). Approximate genome lengths are in Mbases according to http://www.ncbi.nlm.nih.gov/genome/.

| Sample | Accession | Genome length | No. bases | FASTQ files size | Gzipped size | Avg. read length |
|---|---|---|---|---|---|---|
| | | BH1019 family trio | | | | |
| | SRR1778391 | 2,991 | 5.0 | 15.4 | — | 75 |
| | SRR1778392 | 2,991 | 6.3 | 19.4 | — | 75 |
| | SRR1778393 | 2,991 | 6.2 | 19.2 | — | 75 |
| | | AJ family trio | | | | |
| | SRR2962694 | 2,991 | 18.4 | 42.0 | 11.8 | 126 |
| | SRR2962669 | 2,991 | 18.9 | 43.3 | 12.1 | 126 |
| | SRR2962692 | 2,991 | 16.2 | 37.0 | 10.4 | 126 |

The non-public files of BH1019 were downloaded through the dbGaP Authorized Access system and were dumped to paired-end files.

The files of AJ family trio were downloaded from the following URLs:
ftp://ftp.sra.ebi.ac.uk/vol1/fastq/SRR296/004/SRR2962694/
ftp://ftp.sra.ebi.ac.uk/vol1/fastq/SRR296/009/SRR2962669/
ftp://ftp.sra.ebi.ac.uk/vol1/fastq/SRR296/002/SRR2962692/

### 4.1.2 Results

DIAMUND [3] processes sequencing data of family trios. It is designed to detect *de novo* mutations causing disease in a child of healthy parents. In contrast to other methods DIAMUND does not align sequencing reads of family members to a reference genome, which is a resource-expensive task. Instead, it performs a direct comparison of whole genome sequencing (WGS) or whole exome sequencing (WXS) reads of family members. DIAMUND is a ten-step algorithm. A detailed workflow could be found in the original paper. To evaluate our tools we replaced the initial steps of DIAMUND. We also tried to apply GenomeTester4 as a replacement of these steps, but one of them (reads filtering) is not supported by GenomeTester4, hence the pipeline cannot be finished. Nevertheless, the comparison of the earlier steps was performed.

DIAMUND package is distributed with a couple of precomputed files. One of them contains all $k$-mers present in known vectors (such $k$-mers may be present in sequenced sample due to vector contamination). The

other file contains all $k$-mers from a reference exome. These files were converted to KMC and GenomeTester4 database format, separately. The steps that we replaced are as follows (the numbers refer to the original steps in the DIAMUND pipeline).

0. converting the input data to FASTA format and merging paired reads files to a single file. This step is not required in case of application of KMC and GenomeTester4.

1. $k$-mer counting for each family member.

1. converting binary $k$-mer databases to text form and sorting each one. This is necessary only when Jellyfish is used (as in the original DIAMUND) as the further steps are performed on data in a textual format. KMC tools and GenomeTester4 do not require that, because they are able to perform set operations on files in their own binary formats.

2.–4. filtering of set of $k$-mers by removing $k$-mers existing in: both parents, reference exome, and vectors. This requires a few steps in case of original DIAMUND and GenomeTester4 version. In case of KMC tools this may be performed in a single step using *complex* operation.

5. reducing of set of child reads by removing those ones that do not contain any $k$-mer from the filtered child $k$-mer database (being the result of the previous step). DIAMUND uses Kraken algorithm for this step, KMC tools version uses *filter* operation. GenomeTester4 does not support such operation.

We performed two experiments for performance comparison of our improvements with original DIAMUND. First experiment was performed by using one of family trios presented in the DIAMUND paper (BH1019 family trio). The results are presented in Table 9.

Although all experiments presented in the original DIAMUND paper were performed on WXS reads, the authors assure that DIAMUND is also able to handle WGS data. This is a case for our second experiment, for which we use reads from the Ashkenazim Jewish ancestry project (AJ family trio) [4].

As was mentioned earlier, DIAMUND package contains $k$-mers existing in the reference exome and uses them to filter out some of $k$-mers from child. When the input datasets contain WGS reads, $k$-mers existing in the reference genome should be used instead, so we counted all of $k$-mers in the human hg19 genome (also being a part of DIAMUND package). The results of the second experiment are presented in Table 10.

Table 9: Comparison of mutation detection for family trio BH1019. Times are given for: original DIAMUND algorithm, DIAMUND algorithm improved with KMC and KMC tools, DIAMUND algorithm improved with GenomeTester4. The units are: seconds (time), GB (RAM). For all experiments DIAMUND was configured to use 8 threads. '—' means that operation is not required.

| Step | DIAMUND time | KMC and KMC tools time | GenomeTester4 time |
| --- | --- | --- | --- |
| Initial step | 583 | — | — |
| Patient $k$-mer counting | 219 | 156 | 281 |
| Dump and sort patient $k$-mers | 1,597 | — | — |
| Parent 1 $k$-mer counting | 159 | 150 | 278 |
| Dump and sort Parent 1 $k$-mers | 3,375 | — | — |
| Parent 2 $k$-mer counting | 199 | 193 | 330 |
| Dump and sort Parent 2 $k$-mers | 4,179 | — | — |
| Remove vector $k$-mers from patient | 102 |  | 12 |
| Remove reference exome $k$-mers from patient | 217 | 38 | 6 |
| Remove parents $k$-mers from patient | 1,011 |  | 9 |
| Reads filtering | 2,134 | 2,310 | not supported |
| Rest of pipeline | 3,928 | 3,811 |  |
| Replaced steps | 13,775 | 2,847 |  |
| Whole pipeline | 17,703 | 6,658 |  |
| RAM [GB] | 105 | 7 | 41 |

Table 10: Comparison of mutation detection for AJ family trio. Times are given for: original DIAMUND algorithm, DIAMUND algorithm improved with KMC and KMC tools, DIAMUND algorithm improved with GenomeTester4. The units are: seconds (time), GB (RAM). For all experiments DIAMUND was configured to use 8 threads. '—' means that operation is not required.

| Step | DIAMUND time | KMC and KMC tools time | GenomeTester4 time |
| --- | --- | --- | --- |
| Initial step | 1,127 | — | — |
| Patient $k$-mer counting | 830 | 208 | 812 |
| Dump and sort patient $k$-mers | 3,267 | — | — |
| Parent 1 $k$-mer counting | 784 | 211 | 859 |
| Dump and sort Parent 1 $k$-mers | 10,848 | — | — |
| Parent 2 $k$-mer counting | 700 | 180 | 750 |
| Dump and sort Parent 2 $k$-mers | 7,934 | — | — |
| Remove vector $k$-mers from patient | 238 |  | 18 |
| Remove reference genome $k$-mers from patient | 4,731 | 112 | 147 |
| Remove parents $k$-mers from patient | 2,236 |  | 43 |
| Reads filtering | 1,028 | 570 | no supported |
| Rest of pipeline | 12,178 | 12,184 |  |
| Replaced steps | 33,723 | 1,281 |  |
| Whole pipeline | 45,901 | 13,465 |  |
| RAM | 107 | 12 | 76 |

## 4.2 NIKS

### 4.2.1 Datasets

Table 11: Datasets used in the experiments. No. of bases are in Gbases. File sizes are in Gbytes (1Gbyte = $10^9$ bytes). Approximate genome lengths are in Mbases according to http://www.ncbi.nlm.nih.gov/genome/.

| Sample | Accession | Genome length | No. bases | FASTQ files size | Gzipped size | Avg. read length |
|---|---|---|---|---|---|---|
| *Oryza sativa* | | | | | | |
| Hit5500-sd | DRR001789 | 367 | 5.2 | 15.5 | 4.8 | 75 |
| Hit5814-sd | DRR001790 | 367 | 5.6 | 16.1 | 5.2 | 75 |
| Hit1917-sd | DRR001787 | 367 | 6.7 | 19.2 | 6.4 | 75 |
| Hit5243-sm | DRR001791 | 367 | 5.8 | 16.5 | 5.2 | 75 |
| Hit0746-sd | DRR001788 | 367 | 5.6 | 16.0 | 5.1 | 75 |

The files were downloaded from the following URLs:
ftp://ftp.sra.ebi.ac.uk/vol1/fastq/DRR001/DRR001789/
ftp://ftp.sra.ebi.ac.uk/vol1/fastq/DRR001/DRR001790/
ftp://ftp.sra.ebi.ac.uk/vol1/fastq/DRR001/DRR001787/
ftp://ftp.sra.ebi.ac.uk/vol1/fastq/DRR001/DRR001791/
ftp://ftp.sra.ebi.ac.uk/vol1/fastq/DRR001/DRR001788/

### 4.2.2 Results

NIKS [2] is a software designed to detect mutations by a direct comparison of two highly related WGS samples. One of the experiments presented in NIKS paper was designed to find mutations in seven rice mutants by comparing each sample with the other ones pair by pair. We chose five of these rice samples to compare original NIKS with the version improved by us. The goal was to compare the steps that the improvements impact on, so we interrupted the execution of the workflow after those steps. The steps we replaced are:

- $k$-mer counting of both samples — original NIKS uses Jellyfish for this step, our version is based on KMC3,

- construction of a histogram of $k$-mer occurrences to determine threshold of erroneous $k$-mers — original NIKS uses Jellyfish *histo* command, our version uses KMC tools *histogram* command,

- computing $k$-mers specific to each sample — original NIKS uses its own program implemented in Java. This program does not perform strict difference between $k$-mer sets. Instead it produces a set of triples (each consisting of $k$-mer and counters of $k$-mers of both input sets). A specific $k$-mer is present in the output set if:

  - it exists in the first input set and is absent in the second one (the first counter field is taken from the first input, the second counter field is set to zero) or,
  - it exists in both input sets, but the counter of the second one is lower or equal to the previously determined threshold for the second input (counter fields are taken from input sets respectively).

For example, if the input sets of $k$-mers with counters are: $A = \{(AAA, 50), (AAC, 20), (AAT, 15)\}$, $B = \{(AAA, 10), (AAC, 2), (AAG, 20)\}$ and $B_{\text{threshold}} = 5$ then the result of subtraction between $A$ and $B$ is $\{(AAC, 20, 2), (AAT, 15, 0)\}$.

Since such an operation is not a part of KMC tools we prepared a simple C++ program using KMC API. This program is included in the script files. As the input it requires KMC databases in a sorted format

so first we need to perform KMC tools *sort* operation. Our program needs to be executed twice to produce unique *k*-mers for both sets, while the original NIKS is able to achieve the same result in a single run. Nevertheless, even if our solution needs more substages for this step it is much faster.

Table 12: Comparison of the first steps of original NIKS algorithm and improved version with KMC and KMC tools. The units are: seconds (time), GB (RAM). NIKS was configured to use 8 threads for each step. '—' means that operation is not required.

|  | *A*=DRR001787, *B*=DRR001788 | | | | *A*=DRR001787, *B*=DRR001789 | | | |
|  | NIKS | | KMC tools | | NIKS | | KMC tools | |
| Step | Time | RAM | Time | RAM | Time | RAM | Time | RAM |
|---|---|---|---|---|---|---|---|---|
| *k*-mer counting *A* | 525 | 92 | 113 | 12 | 526 | 92 | 118 | 12 |
| histogram *A* | 10 | 8 | 12 | 0 | 10 | 8 | 12 | 0 |
| *k*-mer counting *B* | 442 | 92 | 98 | 12 | 416 | 92 | 98 | 12 |
| histogram *B* | 1 | 5 | 7 | 0 | 2 | 4 | 6 | 0 |
| sort *A* | — | — | 30 | 2 | — | — | 30 | 2 |
| sort *B* | — | — | 20 | 2 | — | — | 18 | 2 |
| remove *B* *k*-mers from *A* | 1,798 | 0 | 32 | 0 | 1,730 | 0 | 30 | 0 |
| remove *A* *k*-mers from *B* |  |  | 39 | 0 |  |  | 38 | 0 |
| **Total** | **2,778** | **92** | **350** | **12** | **2,683** | **92** | **351** | **12** |

Table 13: Comparison of the first steps of original NIKS algorithm and improved version with KMC and KMC tools. The units are: seconds (time), GB (RAM). NIKS was configured to use 8 threads for each step. '—' means that operation is not required.

|  | *A*=DRR001787, *B*=DRR001790 | | | | *A*=DRR001787, *B*=DRR001791 | | | |
|  | NIKS | | KMC tools | | NIKS | | KMC tools | |
| Step | Time | RAM | Time | RAM | Time | RAM | Time | RAM |
|---|---|---|---|---|---|---|---|---|
| *k*-mer counting *A* | 518 | 92 | 120 | 12 | 521 | 92 | 113 | 12 |
| histogram *A* | 4 | 8 | 11 | 0 | 2 | 8 | 12 | 0 |
| *k*-mer counting *B* | 438 | 92 | 101 | 12 | 449 | 92 | 103 | 12 |
| histogram *B* | 1 | 5 | 7 | 0 | 2 | 6 | 8 | 0 |
| sort *A* | — | — | 30 | 2 | — | — | 31 | 2 |
| sort *B* | — | — | 20 | 2 | — | — | 23 | 2 |
| remove *B* *k*-mers from *A* | 1,767 | 0 | 31 | 0 | 1,922 | 0 | 31 | 0 |
| remove *A* *k*-mers from *B* |  |  | 39 | 0 |  |  | 52 | 0 |
| **Total** | **2,728** | **92** | **359** | **12** | **2,896** | **92** | **373** | **12** |

Table 14: Comparison of the first steps of original NIKS algorithm and improved version with KMC and KMC tools. The units are: seconds (time), GB (RAM). NIKS was configured to use 8 threads for each step. '—' means that operation is not required.

|  | $A$=DRR001788, $B$=DRR001789 | | | | $A$=DRR001788, $B$=DRR001790 | | | |
|  | NIKS | | KMC tools | | NIKS | | KMC tools | |
| Step | Time | RAM | Time | RAM | Time | RAM | Time | RAM |
| --- | --- | --- | --- | --- | --- | --- | --- | --- |
| $k$-mer counting $A$ | 535 | 92 | 98 | 12 | 647 | 92 | 100 | 12 |
| histogram $A$ | 2 | 5 | 7 | 0 | 2 | 5 | 7 | 0 |
| $k$-mer counting $B$ | 419 | 92 | 94 | 12 | 441 | 92 | 102 | 12 |
| histogram $B$ | 1 | 4 | 6 | 0 | 2 | 5 | 7 | 0 |
| sort $A$ | — | — | 21 | 2 | — | — | 20 | 2 |
| sort $B$ | — | — | 19 | 2 | — | — | 20 | 2 |
| remove $B$ $k$-mers from $A$ | 1,305 | 0 | 29 | 0 | 1,366 | 0 | 31 | 0 |
| remove $A$ $k$-mers from $B$ |  |  | 28 | 0 |  |  | 0 | 0 |
| **Total** | **2,262** | **92** | **302** | **12** | **2,457** | **92** | **286** | **12** |

Table 15: Comparison of the first steps of original NIKS algorithm and improved version with KMC and KMC tools. The units are: seconds (time), GB (RAM). NIKS was configured to use 8 threads for each step. '—' means that operation is not required.

|  | $A$=DRR001789, $B$=DRR001790 | | | | $A$=DRR001789, $B$=DRR001791 | | | |
|  | NIKS | | KMC tools | | NIKS | | KMC tools | |
| Step | Time | RAM | Time | RAM | Time | RAM | Time | RAM |
| --- | --- | --- | --- | --- | --- | --- | --- | --- |
| $k$-mer counting $A$ | 626 | 92 | 91 | 12 | 451 | 92 | 91 | 12 |
| histogram $A$ | 2 | 4 | 7 | 0 | 2 | 4 | 7 | 0 |
| $k$-mer counting $B$ | 440 | 92 | 102 | 12 | 454 | 92 | 103 | 12 |
| histogram $B$ | 2 | 5 | 7 | 0 | 2 | 6 | 8 | 0 |
| sort $A$ | — | — | 19 | 2 | — | — | 19 | 2 |
| sort $B$ | — | — | 20 | 2 | — | — | 23 | 2 |
| remove $B$ $k$-mers from $A$ | 1,302 | 0 | 29 | 0 | 1,454 | 0 | 32 | 0 |
| remove $A$ $k$-mers from $B$ |  |  | 31 | 0 |  |  | 46 | 0 |
| **Total** | **2,371** | **92** | **307** | **12** | **2,363** | **92** | **330** | **12** |

Table 16: Comparison of the first steps of original NIKS algorithm and improved version with KMC and KMC tools. The units are: seconds (time), GB (RAM). NIKS was configured to use 8 threads for each step. '—' means that operation is not required.

|  | $A$=DRR001790, $B$=DRR001791 | | | | $A$=DRR001788, $B$=DRR001791 | | | |
|  | NIKS | | KMC tools | | NIKS | | KMC tools | |
| Step | Time | RAM | Time | RAM | Time | RAM | Time | RAM |
| --- | --- | --- | --- | --- | --- | --- | --- | --- |
| $k$-mer counting $A$ | 468 | 92 | 102 | 12 | 455 | 92 | 98 | 12 |
| histogram $A$ | 1 | 5 | 7 | 0 | 1 | 5 | 8 | 0 |
| $k$-mer counting $B$ | 459 | 92 | 103 | 12 | 455 | 92 | 102 | 12 |
| histogram $B$ | 2 | 6 | 8 | 0 | 2 | 6 | 8 | 0 |
| sort $A$ | — | — | 20 | 2 | — | — | 20 | 2 |
| sort $B$ | — | — | 23 | 2 | — | — | 22 | 2 |
| remove $B$ $k$-mers from $A$ | 1,485 | 0 | 32 | 0 | 1,517 | 0 | 30 | 0 |
| remove $A$ $k$-mers from $B$ | | | 44 | 0 | | | 42 | 0 |
| **Total** | **2,415** | **92** | **339** | **12** | **2,430** | **92** | **331** | **12** |

## 4.3 GenomeTester4

GenomeTester4 [1] is a software package offering similar functionality to KMC and KMC tools (partial functionality comparison is presented in Table 17). To compare both toolsets we chose one of the experiments presented in the original GenomeTester4 paper. The aim of this experiment was to find $k$-mers specific to a given group of bacteria. We rerun steps required to accomplish this task with GenomeTester4 according to the authors' description. We also designed the pipeline to complete this task with our toolset, which was required due to lack of *difference from union* operation in KMC tools. We replaced it with a couple of supported operations achieving as output the same set of $k$-mers. As the results (Table 18) show, the operations on $k$-mer databases performed by both tools have comparable run times, however, KMC tools require much less memory.

The pipelines are as follows. First we count all $k$-mers of all bacteria to a single database using KMC3. Then we optimize the database using *sort* operation. This is not required, however it allows to perform further operations faster. GenomeTester4 performs similar stage, but firstly all $k$-mers of each bacteria are counted separately using glistmaker, then tens of hundreds result databases are merged pair by pair using *union* operation. For further explanation let we denote the database of all bacteria $k$-mers as $A$. To find $k$-mers specific to a given group we count $k$-mers for each individual and then intersect all of them using *complex* operation (counter calculation mode is set to *sum*). The result of intersection is a database containing $k$-mers occurring in each sample of the group. If a group contains only one sample the intersection is skipped. GenomeTester4 performs similar step but the intersection is made in pairs. Let we denote the database containing the result of intersection as $B$. All equal $k$-mers with equal counters in $A$ and $B$ are specific to the group of interest. GenomeTester4 uses operation dedicated to such situation, in case of KMC tools further computations are necessary. Intersection of $A$ and $B$ with counter calculation mode set to *max* is performed, the result is stored in $C$. The *counters_subtract* operation is performed on $C$ and $B$ and the result is stored in $D$ (it contains $k$-mers that exist not only in the group of interest and should be removed from $B$). The last operation is *kmers_subtract* performed on $B$ and $D$ which result are $k$-mers specific to the given group.

Table 17: Comparison of functionalities offered by GenomeTester4 and KMC tools

| Functionality | GenomeTester4 | KMC tools |
|---|---|---|
| Supported $k$ | $\leq 32$ | any |
| Redefinition of counter calculation mode | possible | possible |
| Maximal no. of output databases per single run | couple (one per operation) | any |
| No. of input databases | any for union, two for other operations | any |
| Mismatch support | yes (only for difference) | no |
| Support for counter-aware subtraction | no | yes |
| Support for counter-aware intersection | yes | no |
| Supprot for user-defined-operation expressions | no | yes |

Table 18: Comparison of GenomeTester4 and KMC tools on finding group specific $k$-mers. The units are: seconds (time), GB (RAM)

| KMC tools | | | GenomeTester4 | | |
|---|---|---|---|---|---|
| Step | RAM | Time | Step | RAM | Time |
| **Counting $k$-mers of all bacteria** | | | | | |
| $k$-mer counting | 12.0 | 468.8 | $k$-mer counting | 0.7 | 4,484.2 |
| database optimization | 3.8 | 154.9 | union of all databases | 75.0 | 13,567.8 |
| **Total:** | **12.0** | **623.7** | | **75.0** | **18,052.0** |
| **Finding *Streptococcus* specific $k$-mers** | | | | | |
| $k$-mer counting | 0.0 | 41.2 | | | |
| intersection | 2.4 | 16.5 | intersection | 0.1 | 14.5 |
| intersection 2 | 0.2 | 53.0 | | | |
| counters difference | 0.0 | 0.2 | | | |
| $k$-mers difference | 0.0 | 0.1 | difference from union | 68.1 | 36.7 |
| **Total:** | **2.4** | **111.0** | | **68.1** | **51.2** |
| **Finding *S. pneumoniae* specific $k$-mers** | | | | | |
| $k$-mer counting | 0.0 | 16.0 | | | |
| intersection | 0.6 | 1.2 | intersection | 0.1 | 7.4 |
| intersection 2 | 0.2 | 54.9 | | | |
| counters difference | 0.1 | 0.3 | | | |
| $k$-mers difference | 0.0 | 0.2 | difference from union | 68.1 | 37.5 |
| **Total:** | **0.6** | **72.6** | | **68.1** | **44.9** |
| **Finding *S. pneumoniae* G54 specific $k$-mers** | | | | | |
| $k$-mer counting | 0.0 | 0.6 | | | |
| intersection 2 | 0.2 | 69.2 | | | |
| counters difference | 0.1 | 0.5 | | | |
| $k$-mers difference | 0.1 | 0.3 | difference from union | 68.1 | 120.6 |
| **Total:** | **0.2** | **70.6** | | **68.1** | **165.5** |
| **Summary:** | **12.0** | **877.9** | | **75.0** | **18,313.6** |



\end{document}